\documentclass[preprint,nofootinbib]{revtex4}
\usepackage{graphicx}
\usepackage{amsfonts,epsf}
\usepackage{amssymb,epsfig}
\usepackage{latexsym}
\usepackage{amsmath}
\usepackage{tabularx}

\usepackage{multirow}
\usepackage{bm}
\usepackage{ifsym}

\usepackage{threeparttable}

\usepackage{natbib}
\usepackage[bf, sf, FIGTOPCAP, nooneline, tight]{subfigure}
\usepackage{bm}
\usepackage[usenames,dvipsnames]{color}
\usepackage[colorlinks=true, linkcolor=BrickRed, citecolor=Blue, urlcolor=Blue, filecolor=Blue]{hyperref}
\usepackage{booktabs}
\usepackage{hhline}
\usepackage{epsf}
\usepackage{color}
\usepackage{epsf}
\usepackage{psfrag}
\usepackage{epsfig}
\usepackage{epstopdf}

\RequirePackage{color}

\begin{document}
\newcommand{\OM}{\Omega_M}
\newcommand{\Om}{\Omega_m}
\newcommand{\Omo}{\Omega_m^{(0)}}
\newcommand{\Omh}{\hat{\Omega}_m}
\newcommand{\OMo}{\Omega_{M}^0}
\newcommand{\Oro}{\Omega_{r}^0}
\newcommand{\OL}{\Omega_{\Lambda}}
\newcommand{\Oc}{\Omega_{c}}
\newcommand{\Oco}{\Omega_{c}^{(0)}}
\newcommand{\OLh}{\hat{\Omega}_{\Lambda}}
\newcommand{\GL}{\Gamma_{\Lambda}}
\newcommand{\GLh}{\hat{\Gamma}_{\Lambda}}
\newcommand{\OLo}{\Omega_{\Lambda}^0}
\newcommand{\OX}{\Omega_{X}}
\newcommand{\OXo}{\Omega_{X}^0}
\newcommand{\OXh}{\hat{\Omega}_{X}}
\newcommand{\OD}{\Omega_{D}}
\newcommand{\ODo}{\Omega_{D}^{(0)}}
\newcommand{\OR}{\Omega_R}
\newcommand{\OK}{\Omega_K}
\newcommand{\OKo}{\Omega_{K}^0}
\newcommand{\OZ}{\Omega_0}
\newcommand{\OT}{\Omega_T}
\newcommand{\rc}{\rho_c}
\newcommand{\rco}{\rho_{c}^{(0)}}
\newcommand{\rmo}{\rho_{m}^{(0)}}
\newcommand{\rro}{\rho_{r}^0}
\newcommand{\rs}{\rho_s}
\newcommand{\ps}{p_s}
\newcommand{\rM}{\rho_M}
\newcommand{\rmr}{\rho_m}
\newcommand{\pmr}{p_m}
\newcommand{\rMo}{\rho_{M}^0}
\newcommand{\pM}{p_M}
\newcommand{\rR}{\rho_r}
\newcommand{\rD}{\rho_D}
\newcommand{\rDt}{\tilde{\rho}_D}
\newcommand{\rDo}{\rho_{D}^{(0)}}
\newcommand{\rX}{\rho_X}
\newcommand{\pX}{p_X}
\newcommand{\wX}{\omega_X}
\newcommand{\wm}{\omega_m}
\newcommand{\wCC}{\omega_\CC}
\newcommand{\wR}{\omega_R}
\newcommand{\aR}{\alpha_R}
\newcommand{\amr}{\alpha_m}
\newcommand{\aef}{\alpha_e}
\newcommand{\aX}{\alpha_X}
\newcommand{\rL}{\rho_{\CC}}
\newcommand{\pL}{p_{\CC}}
\newcommand{\rLo}{\rho_{\CC}^0}
\newcommand{\pD}{p_D}
\newcommand{\wD}{\omega_D}
\newcommand{\wDo}{\omega_D^{(0)}}
\newcommand{\zm}{z_{\rm max}}
\newcommand{\wL}{\omega_{\CC}}
\newcommand{\CC}{\Lambda}
\newcommand{\CCo}{\Lambda_0}
\newcommand{\bCC}{\beta_{\Lambda}}
\newcommand{\we}{\omega_{e}}
\newcommand{\re}{r_{\epsilon}}
\newcommand{\tOM}{\tilde{\Omega}_M}
\newcommand{\tOm}{\tilde{\Omega}_m}
\newcommand{\tOmo}{\tilde{\Omega}_m^0}
\newcommand{\tOL}{\tilde{\Omega}_{\CC}}
\newcommand{\tOD}{\tilde{\Omega}_{D}}
\newcommand{\tODo}{\tilde{\Omega}_{D}^0}
\newcommand{\xL}{\xi_{\CC}}
\newcommand{\fM}{f_{M}}
\newcommand{\fL}{f_{\Lambda}}
\newcommand{\lu}{\lambda_1}
\newcommand{\ld}{\lambda_2}
\newcommand{\lt}{\lambda_3}
\newcommand{\model}{X$\CC$CDM}
\newcommand{\f}{\tilde{f}}
\newcommand{\cM}{{\cal M}}
\newcommand{\cMd}{{\cal M}^2}
\newcommand{\ka}{\kappa}
\newcommand{\xiu}{\zeta}
\newcommand{\xiR}{\xi'}
\newcommand{\prm}{\delta\rho_m}
\newcommand{\prD}{\delta\rho_D}
\newcommand{\poD}{\delta\omega_D}

\newcommand{\CH}{C_H}
\newcommand{\CHd}{C_{\dot{H}}}
\newcommand{\Hd}{\dot{H}}
\newcommand{\rDE}{\rho_{\rm DE}}
\newcommand{\tetm}{\theta_{\rm m}}
\newcommand{\rplu}{r_{+}}
\newcommand{\rmin}{r_{-}}
\newcommand{\nueff}{\nu_{\rm eff}}
\newcommand{\zeff}{\zeta_{\rm eff}}
\newcommand{\xim}{\xi_m}
\newcommand{\rRo}{\rho_r^{(0)}}
\newcommand{\be}{\begin{equation}}
\newcommand{\ee}{\end{equation}}

\newcommand{\DA}{${\cal D}$A}
\newcommand{\DAU}{${\cal D}$A1}
\newcommand{\DAD}{${\cal D}$A2}
\newcommand{\DAT}{${\cal D}$A3}
\newcommand{\DCU}{${\cal D}$C1}
\newcommand{\DCD}{${\cal D}$C2}
\newcommand{\DC}{${\cal D}$C}
\newcommand{\DHlin}{${\cal D}$H}
\title{Hubble-rate-dependent Dark Energy in Brans-Dicke
Cosmology}
\author{E. Karimkhani \footnote{Email:\text{Elahe.karimkhani@gmail.com}}~and~
A.Khodam-Mohammadi \footnote{Email:\text{khodam@basu.ac.ir
}}} \affiliation{Department of Physics,
Faculty of Science, Bu-Ali Sina University, Hamedan 65178, Iran}


\begin{abstract}
Three general cases of dynamical interacting dark energy models
($\mathcal{D}$-class) are investigated in the context of Brans-Dicke
cosmology. Some of important cosmological quantities are calculated for every cases as a function of
redshift parameter. The most important part of this paper deals with
fitting models with two different expansion history:
(SNIa+BAO$_A$+$Omh^{2}$ and SNIa+BAO$_A$+H(z)) and with two different sets of data for Hubble parameter. This provides a remarkable feature to could analytically see the effects of each analyzes and each data sets on final results. The best fitted values of
parameters according to these analyzes and data points, $\chi_{tot}^2/dof$,  AIC and BIC are reported. By these diagnostic tools we found that some of these models have no chance against $\Lambda$CDM, even without need to study the structure formation, and could be ruled out. While some (e.g.
$\mathcal{BD-D}C2$ and $\mathcal{BD-D}A^*$) render the best fit
quality,i.e. the value of AIC and BIC and figures show that they fit perfectly with overall
data and reveals a strong evidence in favor of these two models against
$\Lambda$CDM.
\end{abstract}

\maketitle

\section{ INTRODUCTION}

\label{sec:intro}

The concordance model is one of the famous dark energy models (DE),
which is supported by numerous observations. The subsequent
measurements of distant supernova \citep{Perlmutter:1998np,
Riess:1998cb} and most recently from the analysis of the precision
cosmological data by the Planck collaboration \citep {Ade:2015xua},
reveals an accelerating expanding universe. Despite of good
consistency with measurements, it suffers with two profound
problems. One of them, which is the most theoretical enigmas of
fundamental physics, so-called cosmological constant problem
\citep{Sola:2013gha,Sola:2015csa}, or fine tuning, and the second
one is Cosmic Coincidence problem (see for instance
\citep{Copeland:2006wr, Padmanabhan:2002ji, Peebles:2002gy,
Sola:2013gha} for further information). The former, namely the
preposterous mismatch between the measured value of cosmological
observations and the typical prediction of $\Lambda$ in quantum
field theory (QFT) \citep{Perlmutter:1998np, Riess:1998cb,
Ade:2015xua} and the latter discus about the ratio of dark matter to
dark energy densities which must be bound into order of unity. It is
a matter of fact that whether the cosmological constant, $\Lambda$
or its density of energy ,$\rho_{\Lambda}=\Lambda/8\pi G$, is truly
a constant or instead is a function of time (or scale factor $a(t)$
or Hubble rate $H(t)$). It is important to note that each model must
satisfy at the same time theoretical considerations and
observational evidences. Following this, different scenarios have
been proposed. From one side, recently, a class of dynamical vacuum
dark energy models (DVM's) was introduced \citep{Sola:2015rra} in
which $\Lambda$ can be considered as a function of Hubble rate $H$,
$ \Lambda(H)=n_0+n_1 H^2+n_2 \dot{H}+...$ \citep{Sola:2016vis} with
the equation of state parameter like the CC (i.e. $w=P/\rho=-1$).
Some authors have also considered an interaction between dark matter
and dark energy in framework of the flat
Friedmann-Lema\^{\i}ter-Robertson-Walker (FLRW) in GR and found a
``strong evidence" against the $\Lambda$CDM \citep{Sola:2016ecz} ,
in favor of the DVM's.\newline From the other side, many authors
interested to consider dynamical DE models, with time varying
$w(t)$, such as: scalar fields, both quintessence and phantom-like,
modified gravity theories, phenomenological decaying vacuum energy
models, holography scenarios, and etc. (more detail is referred to
previous review articles, references therein, and also
\citep{Copeland:2006wr, Li:2011sd} ). These models, can however
alleviate the cosmological problems, specially cosmic coincidence
problem, while less investigation on fine tuning
problem can be found in literature. Recently, one of us with others studied the cosmological
implications and linear structure formation of such dynamical dark
energy models, so-called $\mathcal{D}$-class. Where they have shown $\mathcal{D}$-models improve significantly the fit quality of the $%
\Lambda$CDM and besides, a moderate dynamical DE behavior is better
than having a rigid $\Lambda$-term for the entire cosmic history
\citep{Gomez-Valent:2015pia}.

Now we are at the point that using this kind of dynamical DE into
the Brans-Dicke (BD) theory of gravity. This has been
considered as a scalar-tensor theory, introduced by Jordan
\citep{Jordan:1949zz}, based on the
Mach's principle which is a fundamental principle to explain the
origin of the inertia and then ripened by brans and Dicke
\citep{Brans:1961sx, Dicke:1961gz}. In attempting to incorporate the Mach's
principle, the BD theory introduces a time dependent inertial scalar
field $\varphi $, which plays the role of the gravitational constant
$G$, so that $<\varphi (t)>\propto 1/G$ and is determined by the
distribution of mass of the universe. So the gravitational field is
described by the metric $g_{\mu\nu}$ and the BD scalar field
$\varphi $, which has the dimension $[\varphi ]=[M]^{2}$. In BD
theory, the scalar field $\varphi $ couples to gravity via a
coupling parameter $\omega $ and it has been generalized for various
scalar tensor theories. This theory passes the observational tests
in the solar system domains \citep{Bertotti:2003rm} and also has
been examined by some famous cosmological tests such as Cosmic
Microwave Background (CMB) and Large Scale Structure (LSS)
\citep{Chen:1999qh, Acquaviva:2007mm, Tsujikawa:2008uc, Wu:2009zb,
Li:2013nwa, Javier:2018sol}. In recent years, many authors have been studied on the
some models of DE (e.g. Holographic DE, Ricci DE, Ghost DE, and
etc.) in the BD cosmology and have been found good result and
fitting with observational data. Most of these models can fit in the
category of general $\mathcal{D}$-class DE models. Besides, in an interesting investigation \citep{sola-karimkhani:2016, sola:2018-BD} general time variation of fundamental constants in the context of BD theory is predicted where  new clues for solving CC problem is provided. Hence, this could be a
good motivation for assay this class of DE models in the context of the BD theory to check if it will reveal better analyze than rigid $\Lambda CDM$ model or not.

This paper is organized as follows. After a brief review on the
Brans-Dicke cosmology, we introduce three classes of dynamical DE in
Sec. \ref{sect:BD idea}. The background solution and cosmological
implications of each class of DE models are studied by different
subsections in Sec \ref{sect:CBS}. The fitting of models by the
observational data and make constraint of parameters in each case
are performed in Sec. \ref{sect:Fitting}. At the following, in
Sec. \ref{sec:results}, we give a detailed discussion on the results
by studying on the best fit quality and the chance of each case in
the competition of $\Lambda$CDM. At last, we finished our paper by
some concluding and remarks.

\section{General Formalism: Dynamical DE in the framework of BD cosmology}

\label{sect:BD idea} The BD action has been given by
\begin{equation}
s=\int d^{4}x\sqrt{-g}\left( \phi R-\frac{\omega }{\phi }\partial _{\mu
}\phi \partial ^{\mu }\phi +\mathcal{L}_{m}\right) ,  \label{eq:BD action}
\end{equation}%
where $\phi $ is the BD scalar field, $\omega $ is the BD coupling
parameter and $\mathcal{L}_{m}$ is the Lagrangian of the
pressureless cold dark matter (CDM). General relativity is a
particular case of the BD theory, corresponding to $\omega
\rightarrow \infty $ \citep{Weinberg:1988cp}. In a flat FRW
universe, the BD field equations in a natural unit have been given
by \citep{Banerjee:2000gt}
\begin{eqnarray}
&&3H^{2}-\frac{1}{2}\omega \frac{\dot{\phi}^{2}}{\phi ^{2}}+3H\frac{\dot{\phi}%
}{\phi } =\frac{1}{\phi }(\rho _{m}+\rho _{D})  \label{eq:eom-1} \\
&&2\dot{H}+3H^{2}+\frac{1}{2}\omega \frac{\dot{\phi}^{2}}{\phi ^{2}}+2H\frac{%
\dot{\phi}}{\phi }+\frac{\ddot{\phi}}{\phi } =-\frac{1}{\phi
}p_{D}, \label{eq:eom-2}\\
&&\ddot{\phi}+3H\dot{\phi}=\frac{1}{2\omega
+3}(\rho_m+\rho_D-3P_D)\label{kelain}
\end{eqnarray}%
where $H=\dot{a}/a$ is the Hubble rate and the over dot denotes a
derivative with respect to the cosmic time. At the following, we are
interested to consider that the total energy contents of our
universe including a pressureless CDM, a DE fluid which
its dynamical equation of state (EoS) defines as
$w_{D}={p_{D}}/{\rho _{D}}$ and ignoring any radiation component.
\newline As it is common in literatures, we also assume that the BD
scalar field is proportional to the scale factor: $\phi =\phi
_{0}a^{n}=\phi _{0}(1+z)^{-n}$ where $z$ is redshift and $n$ possess
a tiny value in order to have a slowly time varying of $G$, which is
consist with our foundation about the universe. It is worthwhile to
mention that $n$ will be considered as a free parameter and must be
fitted by the observational data. By inserting scale factor
dependence of $\phi $ in Eqs. (\ref{eq:eom-1}) and (\ref{eq:eom-2}),
we find
\begin{eqnarray}
\rho _{D} &=&\frac{3\phi H^{2}\varsigma }{(1+u)},  \label{eq:BDrho1} \\
\rho _{D} &=&-\frac{H^{2}\phi }{w_{D}}\left( \frac{\dot{{H}}}{H^{2}}%
(2+n)+\vartheta \right) .  \label{eq:BDrho2}
\end{eqnarray}%
where new parameters $\varsigma =1+n-\omega n^{2}/6$ and $\vartheta
=3+2n+n^{2}+\omega n^{2}/2=-3\varsigma +n^{2}+5n+6$ are constants
and $u=\rho _{m}/\rho _{D}$ defines as the ratio of DE to DM densities.
As one may examin in the limit $n=0$ , the standard Friedmann equations will
be recovered. Let's remark that if we define the critical density at
present time as $^1$\footnotetext[1]{One may defines $\rho
_{c}^{(0)}=3H_{0}^{2}\phi _{0}\varsigma $ and hence
Eq. (\ref{eq:matterdensity}) reduced to $\Omega _{m}^{(0)}=\frac{u_{0}}{%
1+u_{0}}$ which is fixed for the present time with no dependence to
free parameters of models that will be explained in Secs.
\ref{sect:BDDA1 model}, \ref{sect:BDDC1 model}, \ref{sect:BDDC2
model} and thus it is not preferred here.} $\rho
_{c}^{(0)}=3H_{0}^{2}\phi _{0}$, then the Friedmann equation (\ref {eq:eom-1}) yields
\begin{equation}
\Omega _{m}^{(0)}+\Omega _{D}^{(0)}+\Omega _{\phi }^{(0)}=1.
\label{eq:density parameter}
\end{equation}%
Here $\Omega _{\phi }^{(0)}=\frac{1}{6}\omega n^{2}-n$ and the matter density
parameter at present time take the following simple form
\begin{equation}
\Omega _{m}^{(0)}=\frac{u_{0}}{1+u_{0}}\varsigma ,  \label{eq:matterdensity}
\end{equation}%
where $u_{0}$ is the value of energy density ratio at present. At some
points in next sections ( \ref{sect:BDDA1 model}, \ref{sect:BDDC1 model} and %
\ref{sect:BDDC2 model}), we will show that in order to determine the
evolution of energy density with respect to redshift $z$, we must
fix $u_{0}$ and accordingly $\Omega _{m}^{(0)}$ parameter at
present. But, as one may find from Eq. (\ref{eq:matterdensity}),
these two parameters will be related to observation due to free
parameter $n$, which is hidden in the parameter $\varsigma
$.\newline Considering Eqs. (\ref{eq:BDrho1}) and (\ref{eq:BDrho2}),
we will gain a general equations which will be beneficial for our
purpose in next sections as:
\begin{equation}
\frac{\dot{H}}{H^{2}}=\frac{-3w_{D}\varsigma -\vartheta (1+u)}{(2+n)(1+u)}.
\label{eq:DHH22-general}
\end{equation}%
The DE density and its dynamical nature plays a crucial role on the
evolution of the universe. At the following we will consider three
basic cases of Hubble-rate-dependent dynamical DE models as
\begin{eqnarray}
\mathcal{BD-D}A1:\phantom{XX}\rho _{D}(H) &=&{3}\phi \left( \alpha
H^{2}+\epsilon \right) ,  \notag  \label{eq:ModelsA} \\
\mathcal{BD-D}C1:\phantom{XX}\rho _{D}(H) &=&{3}\phi (\alpha H^{2}+\beta H),
\notag \\
\mathcal{BD-D}C2:\phantom{XX}\rho _{D}(H) &=&{3}\phi (\alpha H^{2}+\gamma
\dot{H}). \label{eq:ModelsC}
\end{eqnarray}%
Note that $%
\phi $ has dimension 2 (mass square) and two parameters $\alpha $,
$\gamma $ are dimensionless but two $\beta $, $\epsilon $ have dimensions
1 and 2 in turn. Free parameters $\alpha $ and $\gamma $ will be
fitted by the observational data while $\beta $ and $\epsilon $
can be restricted and related to other free parameters of each case.\\
Another point is that these different DE densities definitions, introduced in Eq. (\ref{eq:ModelsC}), have not been derived from variation of BD action, Eq. (\ref{eq:BD action}). The philosophy of such definitions is as what has been explained in \citep{Gomez-Valent:2015pia}, but by this difference that here, in the context of BD theory, we have used $\phi=\frac{1}{8\pi G}$, see also \citep{sola-karimkhani:2016}.
 
\section{Cosmological background solution}

\label{sect:CBS} At the following, assuming two dark components (DE and DM) for cosmic fluid, we will
consider two scenarios:\newline
$\dagger$) Interacting model: In this case, two components do not conserve separately
and interact with each other in such a manner that the continuity equation for each components
take the form
\begin{equation}
\dot{\rho}_{D}+3H(1+w_{D})\rho _{D}=-Q,  \label{eq:continde}
\end{equation}
\begin{equation}
\dot{\rho}_{m}+3H\rho _{m}=Q,  \label{eq:contindm}
\end{equation}
where $Q$ stands for the interaction term. The idea of this type of interaction has been motivated by     the theory of quantum gravity but it has been chosen by a pure dimensional basis up to now. Usually in litterateurs,
the interaction term is defined in any of the following forms: (i) $Q\propto
H \rho_D$, (ii) $Q\propto H \rho_m$, or (iii) $Q\propto H (\rho_m+\rho_D)$.
Thus hereafter we choose only the first case, $Q=b^2 H \rho_D=\Gamma
\rho _{D}$, where $b^2$ is a free coupling constant parameter.\newline
$\ddagger$) Non-interacting model: In this case two components DM and DE
are considered as self-conserved with no interaction with each other. Then for
obtaining the corresponding equations in this case, it is enough to
substitute $b^2=0$ in all gained equations of first scenario.\newline
By considering the interaction model, from Eqs. (\ref{eq:continde}) and (\ref{eq:contindm}) the evolution of the ratio of energy density can be derived as
\begin{equation}  \label{eq:dudt-general}
\dot{u}=3Hu\left[w_D+\frac{b^2}{3}\left(\frac{1+u}{u}\right)\right].
\end{equation}
Equivalently, changing the cosmic time variable into the redshift due to
relation $d/dt=-(1+z)H(z)d/dz$, leads to
\begin{equation}  \label{eq:dudz-general}
u^\prime(z)=-\frac{3u(z)}{1+z}\left[w_D(z)+\frac{b^2}{3}\left(\frac{1+u(z)}{%
u(z)}\right)\right],
\end{equation}
where prime denotes for derivative with respect to redshift
parameter. Also, for doing a further analysis of background
evolution of the universe, it will be beneficial to calculate
deceleration parameter which is calculated as
\begin{equation}  \label{eq:qz}
q(z)=-1-\frac{\dot{H}}{2H^2}=-1+\frac{1+z}{2H^2}\frac{dH^2}{dz}\,
\end{equation}

\subsection{$\mathcal{BD-D}A1$ case}

\label{sect:BDDA1 model} For $\rho_D=3\phi(\alpha H^2+\epsilon)$ ,
using Eq.(\ref{eq:BDrho1}), the Hubble rate can be given by
\begin{equation}
\ H(t)=\pm \sqrt{\frac{-\epsilon(1+u(t))}{\alpha(1+u(t))-\varsigma}}.
\label{eq:hubble BDDA1}
\end{equation}%
The constant parameter $\epsilon$ can be obtained in terms
of some other    constants by solving eq. (\ref{eq:hubble BDDA1}) at
present time,
\begin{equation}  \label{eq:epsilon}
\epsilon=-H_{0}^2(\alpha-\frac{\varsigma }{1+u_{0}}),
\end{equation}
and the time derivative of Eq. (\ref{eq:hubble BDDA1}) gives
\begin{equation}  \label{eq:BDDA1-DHH21}
\frac{\dot{H}(t)}{H(t)^2}=\frac{\varsigma \dot{u}(t)}{2\sqrt{%
-\epsilon(\alpha(1+u(t))-\varsigma)(1+u(t))^3}}.
\end{equation}
Using Eqs. (\ref{eq:DHH22-general}) and (\ref{eq:BDDA1-DHH21}) and
after changing the parameter $t\rightarrow z$, the EoS parameter can
be given by
\begin{equation}  \label{eq:BDDA1-Eos1}
w_D=-\frac{(1+z)(2+n)u^\prime}{6\left[\alpha(1+u)-\varsigma\right]}-%
\frac{\vartheta(1+u)}{3\varsigma},
\end{equation}
and substituting Eq. (\ref{eq:dudz-general}) in Eq. (\ref{eq:BDDA1-Eos1})
yields
\begin{equation}  \label{eq:BDDA1-udifrential}
u^\prime=\frac{(1+u)}{1+z}\left[\frac{(\frac{\vartheta}{\varsigma}%
u-b^2)}{1-\frac{(2+n)u}{2(\alpha(1+u)-\varsigma)}}\right].
\end{equation}
By solving this equation, the redshift $z$ can be find versus $u$ as
follows
\begin{eqnarray}  \label{eq:BDDA1-u(z)}
z&=&\left[\frac{\vartheta u-\varsigma b^2}{\vartheta u_0-\varsigma b^2}\right]%
^{\frac{\varsigma (2\eta -\varsigma(n+2)b^2)}{2\eta(\vartheta+\varsigma b^2)}%
}\times\left[\frac{1+u}{1+u_0}\right]^{\frac{n-2\varsigma+2}{2(\varsigma
b^2+\vartheta)}}\times \notag \\
&&\left[\frac{\alpha(1+u)-\varsigma}{\alpha(1+u_0)-\varsigma}%
\right]^{\frac{(2+n)(\varsigma-\alpha)}{2\eta}}-1,
\end{eqnarray}
where
\begin{equation}  \label{eq:eta}
\eta=\vartheta(\alpha-\varsigma)+\alpha\varsigma b^2.
\end{equation}
Finally the EoS parameter (\ref{eq:BDDA1-Eos1}) and deceleration parameter (%
\ref{eq:qz}), in term of energy density ratio by using of Eq. (\ref%
{eq:BDDA1-udifrential}), can be rewritten as
\begin{eqnarray}
w_D&=&\left[\frac{%
(2+n)b^2\varsigma-2\vartheta(\alpha(1+u)-\varsigma)}{2(\alpha(1+u)-%
\varsigma)-(2+n)u}\right]\times \notag
\\&&\left(\frac{1+u}{3\varsigma}\right) ,\label{eq:BDDA1-EoS2}
\end{eqnarray}
\begin{equation}
q=-1-\frac{\vartheta u-b^2\varsigma}{2(\alpha(1+u)-%
\varsigma)-(2+n)u}.  \label{eq:BDDA1-q}
\end{equation}
As it is seen, the EoS and deceleration parameters is not dependent
on constant $\epsilon$ even after considering the explicit formula
of $u(z)$. This result is different with
\citep{Gomez-Valent:2015pia}, where the same DE density was
investigated in the framework of Hilbert-Einstein general relativity
which was called $\mathcal{D}A1$ model there.

At end, it is worthwhile to mention that in limiting case, where $\alpha=n=0$, this case tends to the familiar standard $\Lambda$CDM model (i.e. $\rho=const$).
\subsection{$\mathcal{BD-D}C1$ Model}

\label{sect:BDDC1 model} In this model, $\rho _{D}={3}\phi (\alpha
H^{2}+\beta H)$, using Eq. (\ref{eq:BDrho1}), the Hubble rate takes
the form
\begin{equation}
\ H(t)=-\frac{\beta\left(1+u(t)\right)}{\alpha\left(1+u(t)\right)-\varsigma}.
\label{eq:hubble BDDC1}
\end{equation}%
\ By imposing the current value of Hubble function and energy density ratio
in Eq. (\ref{eq:hubble BDDC1}), one may fix the constant $\beta$ as
\begin{equation}  \label{eq:betaBDDC1}
\beta=-H_0\left(\alpha-\frac{ \varsigma}{1+u_{0}}\right),
\end{equation}

and using Eq. (\ref{eq:hubble BDDC1}), we obtain
\begin{equation}  \label{eq:BDDC1-DHH21}
\frac{\dot{H}(t)}{H(t)^2}=\frac{\varsigma \dot{u}(t)}{\beta(1+u(t))^2}.
\end{equation}
After equating two Eqs. (\ref{eq:BDDC1-DHH21}) and (\ref{eq:DHH22-general}), the EoS
parameter can be calculated as
\begin{equation}  \label{eq:BDDC1-Eos1}
w_D=-\frac{(1+z)(2+n)u^\prime}{3\left(\alpha(1+u)-\varsigma\right)}-%
\frac{\vartheta(1+u)}{3\varsigma}.
\end{equation}
As it is seen, $\beta$ plays no role in the EoS parameter
explicitly. Applying Eq. (\ref{eq:dudz-general}) in Eq.
(\ref{eq:BDDC1-Eos1}) leads to
\begin{equation}  \label{eq:BDDC1-udifrential}
u^\prime=\frac{(1+u)}{1+z}\left[\frac{(\frac{\vartheta}{\varsigma}%
u-b^2)}{1-\frac{(2+n)u}{\alpha(1+u)-\varsigma}}\right],
\end{equation}
and solving above differential equation, (\ref{eq:BDDC1-udifrential}), yields
\begin{eqnarray}  \label{eq:BDDC1-u(z)}
z&=&\left[\frac{\vartheta u-\varsigma b^2}{\vartheta u_0-\varsigma b^2}\right]%
^{\frac{\varsigma (\eta
-\varsigma(n+2)b^2)}{\eta(\vartheta+\varsigma b^2)}}\times %
\left[\frac{1+u}{1+u_0}\right]^{\frac{n-\varsigma+2}{(\varsigma
b^2+\vartheta)}}\times \notag\\
&&\left[\frac{\alpha(1+u)-\varsigma}{\alpha(1+u_0)-\varsigma}%
\right]^{\frac{(2+n)(\varsigma-\alpha)}{\eta}}-1.
\end{eqnarray}
Finally, Eq. (\ref{eq:BDDC1-udifrential}) help us to rewritten the
EoS and deceleration parameters in term of energy density ratio as
\begin{eqnarray}  \label{eq:BDDC1-EoS2}
w_D&=&\left[\frac{(2+n)b^2\varsigma-2%
\vartheta(\alpha(1+u)-\varsigma)}{\alpha(1+u)+\varsigma-(2+n)u}%
\right]\times \notag \\
&&\left(\frac{1+u}{3\varsigma}\right),
\end{eqnarray}
\begin{equation}  \label{eq:BDDC1-q}
q=-1+\frac{\vartheta u-b^2\varsigma}{\varsigma+(2+n)u-\alpha(1+u)}.
\end{equation}
It must be mentioned that the non-interacting case is achieved by
substituting $b^2=0$ in all above relations.

\subsection{$\mathcal{BD-D}C2$ Model}

\label{sect:BDDC2 model} In two previous sections, due to the
special form of DE, after doing a straightforward approach, we were
able to find the Hubble rate with respect to the energy density
ratio. Here, in this section, follow
 \citep{Khodam-Mohammadi:2014wla}, substituting the DE density $\rho_D=3\phi(\alpha H^2+\gamma \dot{H}$) in Eq. (\ref%
{eq:BDrho1}), yields
\begin{equation}  \label{eq:BDDC2-DHH21}
\frac{\dot{H}}{H^2}=\frac{\varsigma}{\gamma(1+u)}-\frac{\alpha}{\gamma}.
\end{equation}
Equating above equation with (\ref{eq:DHH22-general}) gives a
relation between the EoS parameter and energy density ration as
follows
\begin{equation}  \label{eq:BDDC2-EoS1}
w_D=\frac{1}{3}\left[A(1+u)-\frac{2+n}{\gamma}\right],
\end{equation}
where
\begin{equation}
A=\frac{1}{\varsigma}\left[\frac{(2+n)\alpha }{\gamma }-\vartheta\right] .
\label{eq:Adef}
\end{equation}

The deceleration parameter could also be calculated by using (\ref%
{eq:BDDC2-DHH21}) as
\begin{equation}
q=-1+\frac{\alpha }{\gamma }- \frac{\varsigma}{(1+u)\gamma } .
\label{eq:BDDC2-q}
\end{equation}
Substituting Eq. (\ref{eq:BDDC2-EoS1}) in (\ref{eq:dudz-general}), and after solving the obtained differential equation, we find
\begin{eqnarray}
u&=&\frac{1}{2\gamma A}\Big{\{} C\tan \Big{[} -\frac{C \ln(1+z)}{2\gamma }%
+ \notag \\&&\arctan\left( \frac{9\gamma A-5n+5\beta
b^{2}-10)}{5C}\right) \Big{]}
\notag \\
&&-\gamma A+2+n-\gamma b^{2} \Big{\}},  \label{eq:u(x)}
\end{eqnarray}%
where the constant parameter $C$ is given by
\begin{equation}
C=\sqrt{4A\gamma (n+2)-\left( \gamma b^{2}-2-A\gamma -n\right) ^{2}}.
\label{eq:cdef}
\end{equation}
Using the continuity equation (\ref{eq:contindm}), the density of dark
matter becomes
\begin{equation}
{\rho _{m}}={\rho _{m}^{0}}(1+z)^{3}\exp [3b^{2}(\mathcal{F}(z)-\mathcal{F}%
(0))],  \label{eq:rhom}
\end{equation}
in which
\begin{eqnarray}  \label{eq:Fdef}
\mathcal{F}(z)&=&\frac{1}{2A}
\Big{\{}\ln(1+z)(A+ b^{2}-\frac{1}{\gamma })+ \notag \\&&\ln \Big{(} 1+\tan \Big{[} -%
\frac{C \ln(1+z)}{2\gamma }+ \\ &&\arctan \left( \frac{9\gamma
A-5n+5\gamma b^{2}-10)}{5C}\right) \Big{]} ^{2}\Big{)}\Big{\}},
\notag
\end{eqnarray}
where $\mathcal{F}(0)$ is the value of $\mathcal{F}(z)$ at present time. Also,
${\rho _{m}^{0}}$ could be obtain by using (\ref{eq:BDrho1}) as
\begin{equation}  \label{eq:rhom0}
{\rho _{m}^{0}}=\frac{3\varsigma u_{0} H_{0}^2\phi_{0}}{1+u_{0}}.
\end{equation}
At last, the Hubble rate is given by
\begin{equation}
H(z)=\sqrt{\frac{\rho_m(z)}{3 \varsigma \phi(z) }\left(\frac{(1+u(z))}{ u(z)}%
\right)}.  \label{eq:Hubblefun}
\end{equation}

\section{model constraint}

\label{sect:Fitting} In this section, we are interested to extract
the combined data from expansion history: SNIa+BAO$_A$+ $Omh^{2}$ (and SNIa+BAO$_A$+ $H(z)$). We have applied both $Omh^{2}$ and $H(z)$ diagnostic in order to provide better comparison between the results. 
Specifically in \citep{Gomez-Valent:2014rxa, Gomez-Valent:2014fda} a
very detailed description of all these cosmological observables is
provided as well as of the fitting procedure. The interested reader
is refereed to these references for more detail (see also
\citep{Basilakos:2009wi,Grande:2011xf}). To get the best fit values
of the relevant parameters, we maximize the likelihood function,
$\mathcal{L}=e^{{{\chi^{2}}_{tot}}/{2}}$, or equivalently minimize
the joint ${\chi^{2}}_{tot}$ function with respect to the elements
(parameters) of $\mathbf{p}$ where
\begin{equation}
{\chi}_{tot}^{2}(\mathbf{p})={\chi}_{SNIa}^{2}+{\chi}_{BAO_{a}}^{2}+{\chi}%
_{omh^{2}}^{2} (or \; {\chi}_{H(z)}^{2}).  \label{eq:chi2}
\end{equation}
To compare the evidence for and against competing models, it is
common to employ various information criteria like, Akaike
Information Criterion (AIC) and the Bayesian Information Criterion
(BIC), which in addition to $\chi^2$, they take into account the
number of free parameters in each model, $n_{fit}$. Also they are
appropriate for the models which we are studying here
($N_{tot}/n_{fit} > 40$) \citep{1974ITAC...19..716A, Sugiura:1978}.
For the Gaussian errors, they define as:
\begin{equation}
AIC=\chi^2_{tot}+2n_{fit};~~~~~BIC=\chi^2_{tot}+n_{fit}\ln N,
\end{equation}
where $N$, is the number of data points. Two statistical tools AIC
and BIC grade two or more models and give in hand the numerical
measure about each model which is preferred. Any interacting and non-interacting models:
"i"=$\mathcal{BD-D}A1$, $\mathcal{BD-D} C1$ and $\mathcal{BD-D}C2$  that has
smaller value of difference with
respect to "j"=$\Lambda$CDM,there is the evidence in favor of the
shorter one \citep%
{1974ITAC...19..716A, Sugiura:1978, Gomez-Valent:2015pia}. Hence for
a pairwise comparison, the conqueror model is one with positive sign for 
 $\Delta AIC=(\Delta_{ij})_{AIC}=AIC_{j}-AIC_{i}$ and $\Delta BIC=(\Delta_{ij})_{BIC}=BIC_{j}-BIC_{i}$, which is an indication supporting "i" models.\\
But it is needed to have the difference
$\Delta_{ij}\geq2$, because otherwise it betokens as consistency
between these two model in competition, while for
$\Delta_{ij}\geq6$, we will have a strong evidence and $\Delta_{ij}\geq10 $  presents very strong evidence for choosing
preferred model. We will use these issues in the next section. Also, executive explanation over $\Delta$AIC and $\Delta$BIC for each model will be provided in sec. \ref{sec:results}. 
\newline Another point which seems necessary to mention here is that,
in order to constraint each model, we have taken the BD parameter
as $\omega=1033$, which is gained from $PlanckTemp + PlanckLens$ at
$99 \%$ confidence level under unrestricted supposition (no initial
value for scalar field is fixed) \citep{Avilez:2013dxa}. Also it is
consistent with what usually handled in literature (e.g. in \citep%
{Chen:1999qh} the authors has found $\omega\simeq1000$ by using the
CMB temperature and polarization anisotropy data. Also see
\citep{Acquaviva:2007mm, Li:2015aug, Alavirad:2014kqa} and reference
therein).\newline In the following we will explain each of SNIa,
BAO$_A$ , $Omh^{2}$ and $H(z)$ analysis in short.

\subsection{SNIa}

We are using the Union 2.1 set of 580 type Ia supernovae of Suzuki
et al. \citep{Suzuki:2011hu} in the following definition
\begin{equation}
{\chi}_{SNIa}^{2}=\displaystyle\sum_{i=1}^{580}\left[\frac{{\mu}_{th}(z_i,p)-%
{\mu}_{obs}(z_i)}{\sigma_i}\right]^2  \label{eq:chisnia}
\end{equation}
in which $z_i$ is the observed redshift for each data point. The observational
modulus distance of SNIa, $\mu_{obs}(z_i)$, at redshift $z_i$ is given by
\begin{equation}
{\mu}_{obs}(z_i)=m(z_i)-M.
\end{equation}
In theoretical point of view the modulus distance define as ${\mu}%
_{th}(z_i,p)=5\log d_L +25 $, in which $d_L(z_i ,p)$ is the luminosity
distance for spatially flat universe,
\begin{equation}
d_L(z_i ,p)=c(1+z)\int\limits_0^z \frac{dz^\prime }{H(z^\prime )}.
\label{eq:lumdis}
\end{equation}
where $c$ is the speed of light. In computing in this stage, we have fixed H0 = 70 km/s/Mpc, following the setting used in the Union
2.1 sample. The remained parameter $\sigma_i$ is defined as corresponding $1\sigma$ uncertainty for each SNIa data point.%
\newline
It is worthy noting that in models with varying $G$, like BD theory,
a correction must be regarded in order to employ the supernovae
data. In \citep{1982ApJ...254....1A, 1982ApJ...253..785A}, authors
predicted on the basis of an analytic model and reasonable
assumptions that the SN Ia maximum luminosity can be expressed in
terms of ejected nickel mass ($L\propto M_{Ni}$), which with a good
approximation is a fixed fraction of the Chandrasekhar mass
($M_{Ni}\propto M_{Ch}\propto G^{-3/2}$) \citep{1993A&A...270..223K,
GomezGomar:1997iv, Branch:2000zm} and thus for the luminosity
distance we will have $L\propto G^{-3/2}$. Using the definition of
absolute magnitude
\begin{equation}
M=-2.5log\frac{L}{L_\odot}+cte,
\end{equation}
the modulus distance relation must be corrected as
\citep{Li:2015aug}
\begin{eqnarray}
\mu(z)&=&\mu_{obs}^{nc}-\frac{15}{4}\log\frac{G}{G_0}\notag\\
&=&\mu_{obs}^{nc}+\frac{15}{4}\log\frac{\phi(z)}{\phi_0}\notag\\
&&=\mu_{obs}^{nc}-\frac{15}{4}n\log{(1+z)}.
\end{eqnarray}
in which we are using $\phi\propto a^n$ in the third relation and
quantity $\mu_{obs}^{nc}$ is the observed distant modulus before
correction.
\subsection{BAO$_A$}

The BAO measurement at the largest redshift H(z = 2.34) taken after \citep%
{Delubac:2014aqe} on the basis of BAO's in the Ly$\alpha $ forest of
BOSS DR11 quasars. The acoustic parameter A(z), which is collected
by Blake et al. in \citep{Blake:2011en}, has been introduced by
Eisenstein as follows \citep{Eisenstein:2005su}:
\begin{equation}
A(z_{i},p)=\frac{\sqrt{\Omega _{m}^{(0)}}}{{E(z_{i})}^{\frac{1}{3}}}\left[
\frac{1}{z_{i}}\int\limits_{0}^{z_{i}}\frac{dz}{E(z)}\right] ^{\frac{2}{3}},
\end{equation}%
where $E(z)={H(z)}/{H_{0}}$ and $z_{i}$ is the redshift at the place
of observable. In this stage we have used the current value of the
Hubble rate given by the Planck Collaboration \citep{Ade:2015xua}, i.e. $%
H0=67.8km/s/Mpc$. The corresponding $\chi ^{2}$-functions for
$BAO_{A}$ analysis are defined as:
\begin{equation}
{\chi }_{BAO_{A}}^{2}=\displaystyle\sum_{i=1}^{6}\left[ \frac{{A}%
_{th}(z_{i},p)-{A}_{obs}(z_{i})}{\sigma _{A,i}}\right] ^{2}
\label{eq:chibao}
\end{equation}%
where the corresponding values of $z_{i}$, $A_{obs}$ and $\sigma
_{A,i}$ can be obtained from table 3 of \citep{Blake:2011en}.

\subsection{$Omh^{2}$}\label{omh2}
We define the following $%
\chi^2_{Omh^2}$ function, to be minimized:

\begin{equation}  \label{xi2Omh2}
\chi^2_{Omh^2}=\sum_{i=1}^{N-1}\sum_{j=i+1}^{N}\left[\frac{%
	Omh^2_{th}(H_i,H_j)-Omh^2_{obs}(H_i,H_j)}{\sigma_{Omh^2\,i,j}}\right]^2\,,
\end{equation}

\noindent where $N$ is the number of points $H(z)$ contained in the
data set, $H_i\equiv H(z_i)$, and $Omh^2(H_i,H_j)$ is the two-point
diagnostic \citep{Sahni:2014ooa},

\begin{equation}  \label{Omz2}
Omh^2(z_2,z_1)\equiv\frac{h^2(z_2)-h^2(z_1)}{(1+z_2)^3-(1+z_1)^3},
\end{equation}

\noindent with $h(z)/h\equiv H(z)/H_0$, and $\sigma_{Omh^2\,i,j}$ is the
uncertainty associated to the observed value $Omh^2_{obs}(H_i,H_j)$ for a
given pair of points $ij$, viz.

\begin{equation}
\sigma^2_{Omh^2\,i,j}=\frac{4\left[h^2(z_i)\sigma^2_{h(z_i)}+h^2(z_j)%
	\sigma^2_{h(z_j)}\right]}{\left[(1+z_i)^3-(1+z_j)^3\right]^2}\,.
\end{equation}
\\
In order to figure out the effect of various $H(z)$ data sets in the final results,
i.e. $\chi^2$, $AIC$ and $BIC$,  we have benefited from two different data sets in $Omh^2$ diagnostic:\\
1) First set is the available measurements of the Hubble rate as collected in \citep{Ding:2015vpa}. These are essentially the data of \citep{Farooq:2013hq}, with the BAO measurement at the largest redshift $H(z= 2.34)$ taken after \citep{Delubac:2014aqe} and contains 29 data points.\\ 
2) The second values are uncorrelated with the BAO data points and are gained by differential-age technique  employed to passively evolving galaxies and collected in Table 3. of \citep{Sola:2017Hdata} which consists of 30 data points.\\

The outcomes of fit procedure for $Omh^2$ diagnostic and for these two sets are represented in Table. \ref{tableFit2} and Table. \ref{tableFit3}
 in turn. More discussions over this issue will be gathered in Sec. \ref{sec:results}.
 
\subsection{$H(z)$}
Here,instead of the correlated $Omh^2(zi, zj)$
diagnostic in $\chi^{2}_{tot}$ we apply
\begin{equation}
\chi^{2}_{\rm H}({\bf p})=\sum_{i=1}^{30} \left[ \frac{ H(z_{i},{\bf
		p})-H_{\rm obs}(z_{i})} {\sigma_{H,i}} \right]^{2}. \label{chi2H}
\end{equation}

One of our goal in this paper is to reveal ineligibly the inequality in results gained by $Omh^2(zi, zj)$ and  $H(z)$ analyzes in $\chi^{2}_{tot}$.
Furthermore, this will help to provide less correlation and also more precise comparison  between the results.\\
Besides, only second set of $H(z)$ data (explained in subsec.\ref{omh2}) is utilized here( more detail concerning these is presented in \ref{sec:results}) and the results are gathered in Table. \ref{tableFit4}. \\

In \citep{Gomez-Valent:2014rxa} and \citep{Gomez-Valent:2014fda},
more detailed explanation of all of these cosmological observable as
well as on the fitting procedure has been elaborated, and therefore
we have left more detail aside of the present works.

\begin{table*}
    \caption{Best fitted values for the expansion history: BAO$_A$+SNIa+Om$h^2$ and with second set of data points on $H(z_i)$ according to \citep{Ding:2015vpa}, see text, sect:\ref{omh2}}
\begin{center}
\resizebox{1\textwidth}{!}{
	\begin{tabular}{| c  |c | c | c | c  | c | c  |c  |}
		\multicolumn{1}{c}{Model}  & \multicolumn{1}{c}{$\alpha$}  &
		\multicolumn{1}{c}{$\Omega_m^{(0)}$/$\gamma$} &
		\multicolumn{1}{c}{$b^2$} & \multicolumn{1}{c}{$n$}
		&\multicolumn{1}{c}{$\chi^2/dof$} & \multicolumn{1}{c}{$\Delta$AIC}&
		\multicolumn{1}{c}{$\Delta$BIC}
		\\\hline
		{\small $\CC$CDM}  & - & {\small$0.275\pm 0.005$}  & - & -  &
		{\small$808.083/991$} & {\small$0$}& {\small$0$}
		\\\hline
		{\small$\mathcal{BD-D}A1$}  & {\small$0.331\pm 0.022$} & {-} &
		{\small $0.373^{+0.020}_{-0.009}$} & {\small$0.009\pm0.025$} &
		{\small$801.531/989$} & {\small$2.552$}& {\small$-7.248$}
		\\\hline
		{\small$\mathcal{BD-D}$C1 }  &
		{\small$-0.300^{+0.011}_{-0.001}$} & {-} &
		{\small$0.287^{+0.048}_{-0.034}$} & {\small$0.020\pm 0.002$} &
		{\small$800.076/989$} & {\small$4.007$}& {\small$-5.793$}
		\\\hline
		{\small $\mathcal{BD-D}$C2}  & {\small$0.765^{+0.027}_{-0.003}$} &
		{\small$0.430^{+0.008}_{-0.020}$} &
		{\small$0.051^{+0.018}_{-0.004}$} &
		{\small$-0.009^{+0.014}_{-0.006}$}  & {\small$791.735/988$} &
		{\small$10.348$}& {\small$-4.352$}
		\\\hline
		{\small$\mathcal{BD-D}A1^{\star}$}  &
		{\small$-0.073^{+0.003}_{-0.001}$} & {-} & {-} &
		{\small$0.014^{+0.001}_{-0.002}$} & {\small$793.485/990$} &
		{\small$12.598$}& {\small$7.698$}
		\\\hline
		{\small$\mathcal{BD-D}C1^{\star}$}  &
		{\small$-0.315^{+0.003}_{-0.006}$} & {-} &  {-} &
		{\small$0.006^{+0.006}_{-0.001}$} & {\small$815.210/990$} &
		{\small$-9.127$}& {\small$-14.027$}
		\\\hline
		{\small $\mathcal{BD-D}C2^{\star}$}  &
		{\small$0.976^{+0.003}_{-0.051}$} &
		{\small$0.614^{+0.040}_{-0.012}$} & {-} &
		{\small$-0.019^{+0.001}_{-0.007}$}  & {\small$831.811/989$} &
		{\small$-27.728$}& {\small$-37.528$}
		\\\hline
	\end{tabular}}
\end{center}
\label{tableFit2}
\begin{scriptsize}
	\begin{tablenotes}
		\item {\textbf{\textit{NOTE:}}  The best-fitting values for the various models
			and their statistical  significance ($\chi^2$-test,  $\Delta$AIC and $\Delta$BIC see Sect. \ref{sect:Fitting}) for both
			interacting and non-interacting ( indicated by $\star$) cases. All
			quantities corresponds to the expansion history of universe i.e.
			(BAO$_A$+SNIa+Om$h^2$). The given values in third column is
			correspond to $\Omega_m^{(0)}$ (resp. $\gamma$) for $\Lambda$CDM
			(resp. $\mathcal{BD-D}C2$) model. Details of the fitting observables
			are given in Sect. \ref{sect:Fitting}. }
	\end{tablenotes}
\end{scriptsize}

\end{table*}

\begin{table*}
    \caption{Best fitted values for the expansion history: BAO$_A$+SNIa+Om$h^2$ and with second set of data points on $H(z_i)$ obtained by differential-age techniques. }
    \begin{center}
        \resizebox{1\textwidth}{!}{
           \begin{tabular}{| c  |c | c | c | c  | c | c  |c  |}
           	\multicolumn{1}{c}{Model}  & \multicolumn{1}{c}{$\alpha$}  &
           	\multicolumn{1}{c}{$\Omega_m^{(0)}$/$\gamma$} &
           	\multicolumn{1}{c}{$b^2$} & \multicolumn{1}{c}{$n$}
           	&\multicolumn{1}{c}{$\chi^2/dof$} & \multicolumn{1}{c}{$\Delta$AIC}&
           	\multicolumn{1}{c}{$\Delta$BIC}
           	\\\hline
           	{\small $\CC$CDM}  & - & {\small$0.296^{+0.022}_{-0.021}$}  & - & -  &
           	{\small$791.287/1020$} & {\small$0$}& {\small$0$}
           	\\\hline
           	{\small$\mathcal{BD-D}A1$}  & {\small$-0.152^ {+0.013}_{-0.022}$} & {-} &
           	{\small $0.056^{+0.010}_{-0.007}$} & {\small$0.025\pm0.015$} &
           	{\small$763.177/1018$} & {\small$24.110$}& {\small$14.254$}
           	\\\hline
           	{\small$\mathcal{BD-D}$C1}  &
           	{\small$-0.491^{+0.009}_{-0.008}$} & {-} &
           	{\small$0.329^{+0.018}_{-0.024}$} & {\small$0.028\pm 0.030$} &
           	{\small$796.182/1018$} & {\small$-8.895$}& {\small$-18.751$}
           	\\\hline
           	{\small $\mathcal{BD-D}$C2}  & {\small$0.698^{+0.007}_{-0.023}$} &
           	{\small$0.414\pm 0.002$} &
           	{\small$0.061^{+0.008}_{-0.014}$} &
           	{\small$-0.002^{+0.016}_{-0.006}$}  & {\small$776.361/1017$} &
           	{\small$8.926$}& {\small$-5.860$}
           	\\\hline
           	{\small$\mathcal{BD-D}A1^{\star}$}  &
           	{\small$-0.153\pm 0.003$} & {-} & {-} &
           	{\small$0.024^{+0.011}_{-0.012}$} & {\small$763.744/1019$} &
           	{\small$25.543$}& {\small$20.615$}
           	\\\hline
           	{\small$\mathcal{BD-D}C1^{\star}$}  &
           	{\small$-0.534^{+0.013}_{-0.009}$} & {-} &  {-} &
           	{\small$0.019^{+0.016}_{-0.011}$} & {\small$802.908/1019$} &
           	{\small$-13.621$}& {\small$-18.549$}
           	\\\hline
           	{\small $\mathcal{BD-D}C2^{\star}$}  &
           	{\small$0.995^{+0.013}_{-0.041}$} &
           	{\small$0.649\pm 0.23 $} & {-} &
           	{\small$0.007^{+0.011}_{-0.017}$}  & {\small$786.570/1018$} &
           	{\small$0.717$}& {\small$-9.139$}
           	\\\hline
           \end{tabular}}
        \end{center}
       \begin{scriptsize}
       	\begin{tablenotes}
       		\item {\textbf{\textit{NOTE:}} Same as in Table. \ref{tableFit2} All
       		quantities corresponds to the expansion history of universe i.e.
       		(BAO$_A$+SNIa+\textbf{Om$h^2$}). \textbf{But} here we have used the second set of $H(z)$ data, see sect. \ref{omh2}.}
       	\end{tablenotes}
       \end{scriptsize}
       \label{tableFit3}
    \end{table*}

\begin{table*}
    \caption{Best fitted values for the expansion history: BAO$_A$+SNIa+H(z)
    and with second set of data points on $H(z_i)$.}
    \begin{center}
        \resizebox{1\textwidth}{!}{
            \begin{tabular}{| c  |c | c | c | c  | c | c  |c  |}
            	\multicolumn{1}{c}{Model}  & \multicolumn{1}{c}{$\alpha$}  &  \multicolumn{1}{c}{$\Omega_m^{(0)}$/$\gamma$} & \multicolumn{1}{c}{$b^2$} & \multicolumn{1}{c}{$n$}  &\multicolumn{1}{c}{$\chi^2/dof$} & \multicolumn{1}{c}{$\Delta$AIC}&
            	\multicolumn{1}{c}{$\Delta$BIC}
            	\\\hline
            	{\small $\CC$CDM}  & - & {\small$0.286\pm 0.007$}  & - & -  & {\small$580.337/615$}  & {\small$0$}& {\small$0$}
            	\\\hline
            	{\small$\mathcal{BD-D}A1$}  & {\small$-0.020_{+0.011}^{-0.012}$} & {-} & {\small $0.019^{+0.013}_{-0.015}$} & {\small$0.011\pm0.004$} & {\small$579.562/613$} & {\small$-3.225$}& {\small$-12.071$}
            	\\\hline
            	{\small$\mathcal{BD-D}$C1}  & {\small$-0.406^{+0.074}_{-0.118}$} & {-} &  {\small$0.180^{+0.083}_{-0.047}$} & {\small$0.015^{+0.002}_{-0.007}$} & {\small$582.110/613$} & {\small$-5.779$}& {\small$-14.619$}
            	\\\hline
            	{\small $\mathcal{BD-D}$C2}  & {\small$0.934^{+0.06}_{-0.006}$} & {\small$0.606^{+0.011}_{-0.086}$} & {\small$0.002^{+0.001}_{-0.004}$} & {\small$0.008^{+0.003}_{-0.02}$}  & {\small$579.296/612$} & {\small$-4.965$}& {\small$-18.228$}
            	\\\hline
            	{\small$\mathcal{BD-D}A1^{\star}$}  & {\small$-0.036^{+0.005}_{-0.07}$} & {-} & {-} & {\small$0.011^{+0.008}_{-0.001}$} & {\small$579.563/614$} &  {\small$-1.232$}& {\small$-5.649$}
            	\\\hline
            	{\small$\mathcal{BD-D}C1^{\star}$}  & {\small$-0.613^{+0.096}_{-0.009}$} & {-} &  {-} & {\small$0.013^{+0.001}_{-0.007}$} & {\small$583.792/614$} &  {\small$-5.461$}& {\small$-9.878$}
            	\\\hline
            	{\small $\mathcal{BD-D}C2^{\star}$}  & {\small$0.952^{+0.009}_{-0.024}$} & {\small$0.568^{+0.025}_{-0.01}$} & {-} & {\small$0.008\pm{0.003}$}  & {\small$579.361/613$} & {\small$-3.03$}& {\small$-11.870$}
            	\\\hline
            \end{tabular}}
        \end{center}
      \begin{scriptsize}
      	\begin{tablenotes}
      		\item{\textbf{\textit{NOTE:}} Same as in Table. \ref{tableFit2}. \textbf{But} here we Have used $H(z)$ diagnostic instead of Om$h^2$, i.e.
      		all quantities corresponds to (BAO$_A$+SNIa+$H(z)$). \textbf{Beside}, $H(z)$ data comes from set 2, see text,Sect \ref{omh2}.}
      	\end{tablenotes}
        \end{scriptsize}
\label{tableFit4}
    \end{table*}

\section{Discussion and Results}\label{sec:results}

In this section we provide further discussion on the results and
calculations which has been done in previous sections. The plots for
EoS, deceleration parameter and energy density ratio will be
illustrated. At the end, we will see which model place in the
more prominent position in competing with the others and has the
most harmony with observations. \newline
\begin{figure*}
    \centering
    \includegraphics[width=\textwidth]{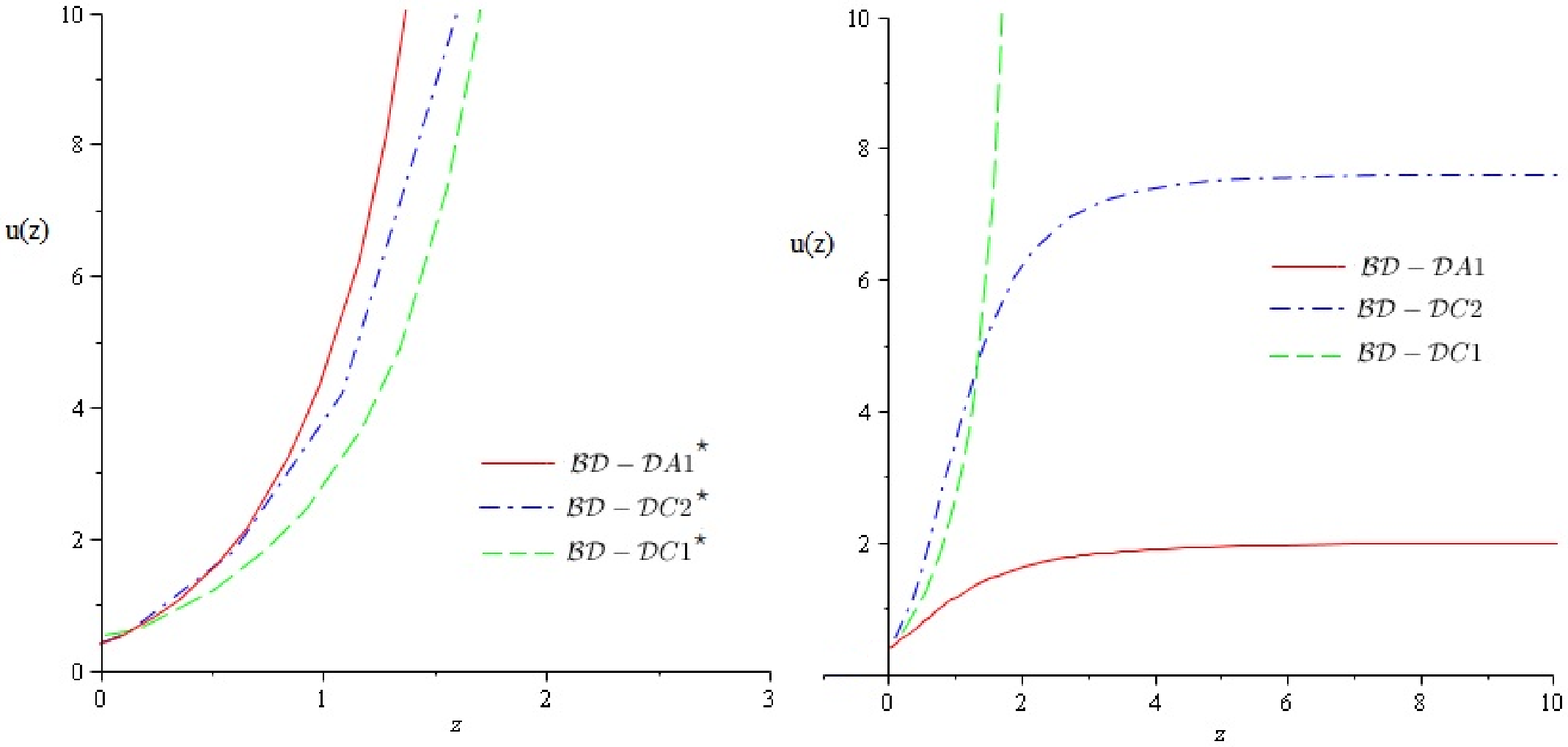}
    \caption{{\protect\footnotesize {\ Energy density ratio versus
                $z$ for non-interacting (left) and interacting (right)  models
                under consideration. Here, we have used the best fit values
                of Table. \ref{tableFit2}}}} \label{fig:u(z)}
\end{figure*}

\begin{figure*}
\centering
    \includegraphics[width=\textwidth]{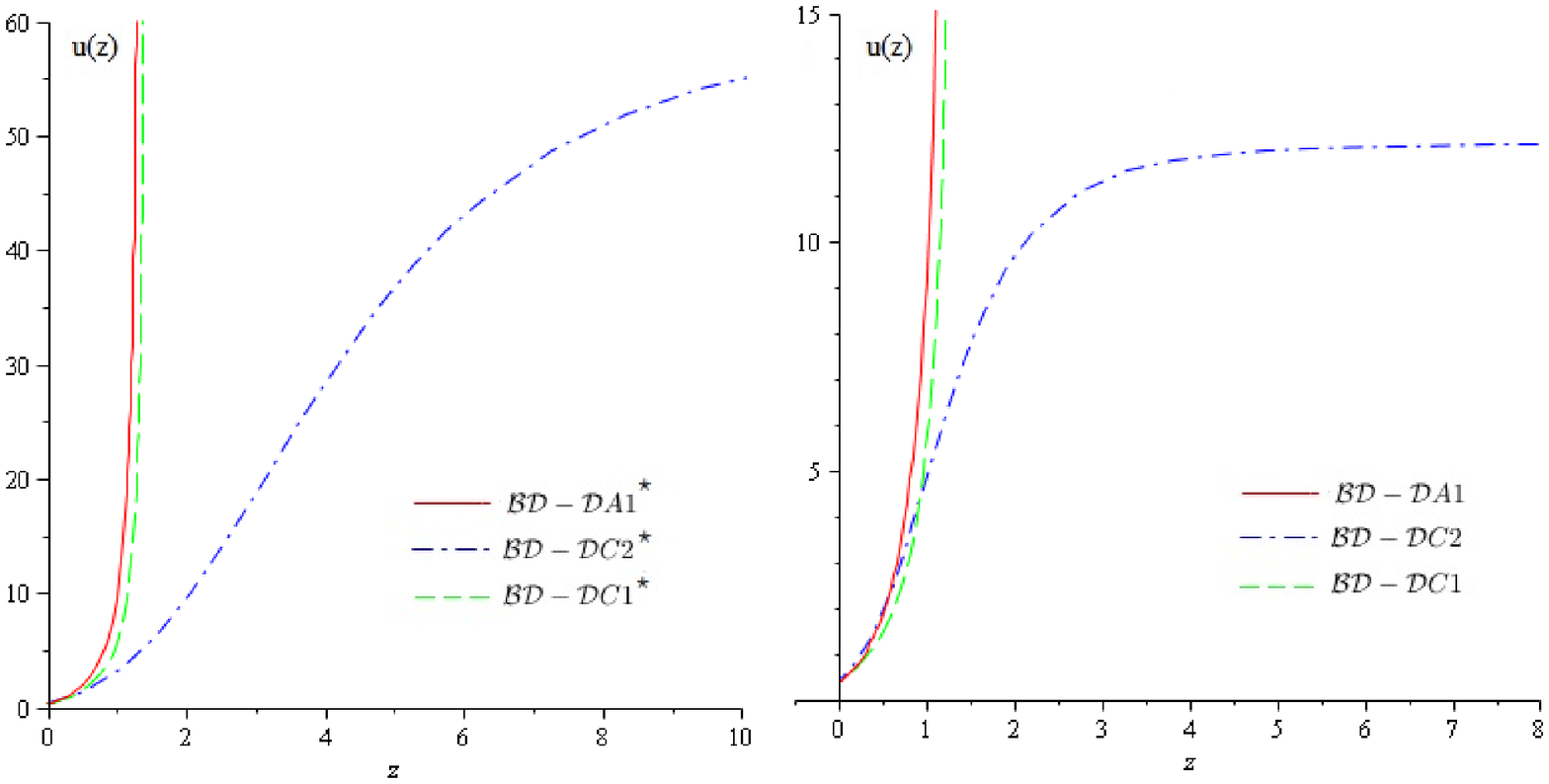}
\caption{{\protect\footnotesize {\ Same as in \ref{fig:u(z)} but here we have used the best fit values of Table. \ref{tableFit3}}}} \label{fig:u(z)-2}
\end{figure*}

\begin{figure*}
\centering
\includegraphics[width=\textwidth]{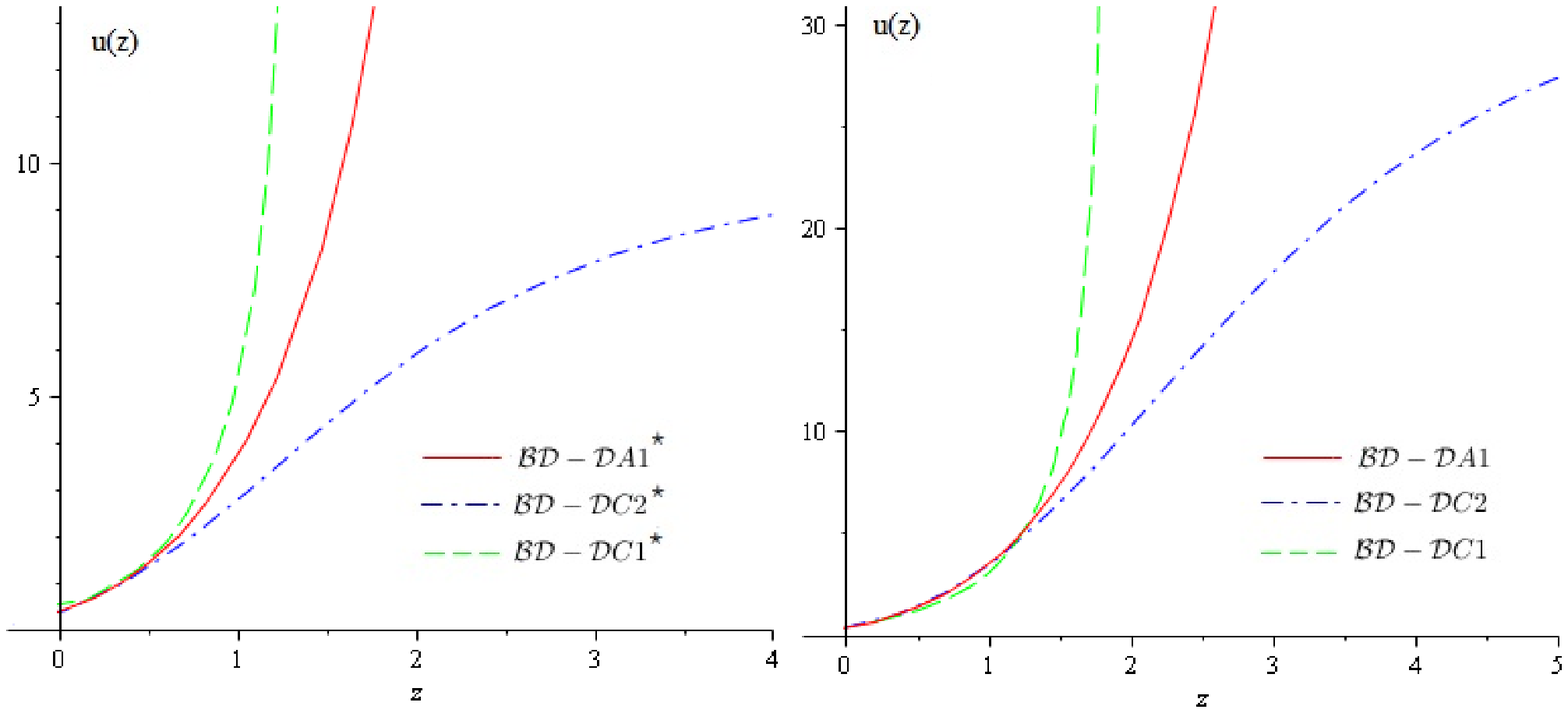}
    \caption{{\protect\footnotesize {\ Same as in \ref{fig:u(z)} but here we  have used the best fit values of Table. \ref{tableFit4}}}} \label{fig:u(z)-3}
\end{figure*}

In tables \ref{tableFit2}, \ref{tableFit3} and \ref{tableFit4}, the best-fitted values of parameters for
each $\mathcal{BD-D}$ models, using the mentioned statistical analysis,
have been collected. These values are used for studying of other
cosmological parameters in the bulk. In these tables
${\chi}_{tot}^{2}/dof$, the $AIC$ and $BIC$ values have been reported in order to appraise the
statistical analyze quality and do better comparison between different cases
studied in this work. \\
The quantity $dof $ is number of
degree of freedom, define as: $dof=N_{tot}-n_{fit}$, where
$N_{tot}$ is total number of data points-dependent on which data sets are
applying and $n_{fit}$ is the
model-dependent number of fitted parameters.\\
As it explained in sec. \ref{sect:Fitting}, in this study we have employed
 two different diagnostics for Hubble parameter: two-point analyze $Omh^{2}$  and $H(z)$.
 We denote these via SNIa+BAO$_A$+$Omh^{2}$ and SNIa+BAO$_A$+$H(z)$ fitting analysis to could show statistically their effects and distinctions on final results.
Also in order to disclose the efficacy of data points in each 
diagnostics we have applied two different data sets for H(z)
parameter which the first set has correlation with BAO and the second are gained with differential age technique.\\
So we have applied both sets of data in SNIa+BAO$_A$+$Omh^{2}$ analyze and
illustrates differences in results causing by each data sets.
Then we have used just uncorrelated data set in SNIa+BAO$_A$+$H(z)$ analyze
and the products are completely  presented in three tables. \ref{tableFit2}, \ref{tableFit3} and \ref{tableFit4}.\\
According to which set of $H(z)$ data and which diagnostics (i.e. $Omh^2$ or $H(z)$ ) we are using, the total number of data points will change. Since for BAO$_A$+SNIa+Om$h^2$ and first set of $H(z)$ data, $N_{tot}=992$, and with second set it will be $N_{tot}=1021$. While for BAO$_A$+SNIa+H(z) and with second data set we have $N_{tot}=616$.\\

\begin{figure*}[tbp]
	\centering
	\includegraphics[width=\textwidth]{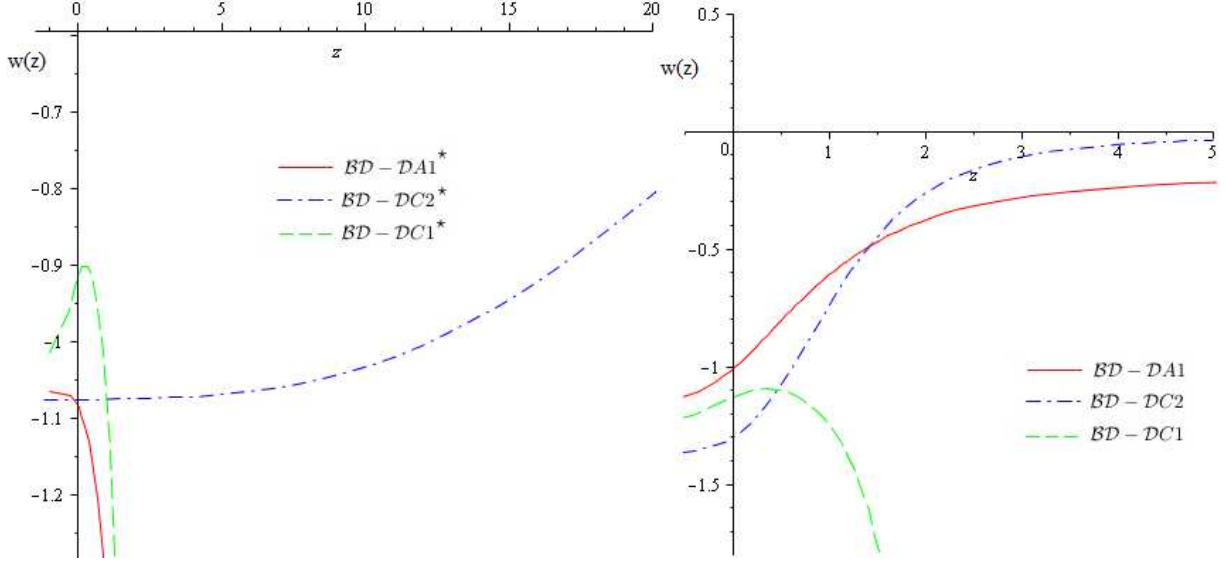}
	\caption{{\protect\footnotesize {\ The evolution of
				$\protect\omega_{D}(z)$ versus $z$, for and non-interacting (left) and interacting
				(right) models using the best fit values
				of Table. \protect\ref{tableFit2}. } }} \label{fig:w(z)}
\end{figure*}

\begin{figure*}
	\centering
	\includegraphics[totalheight=8cm]{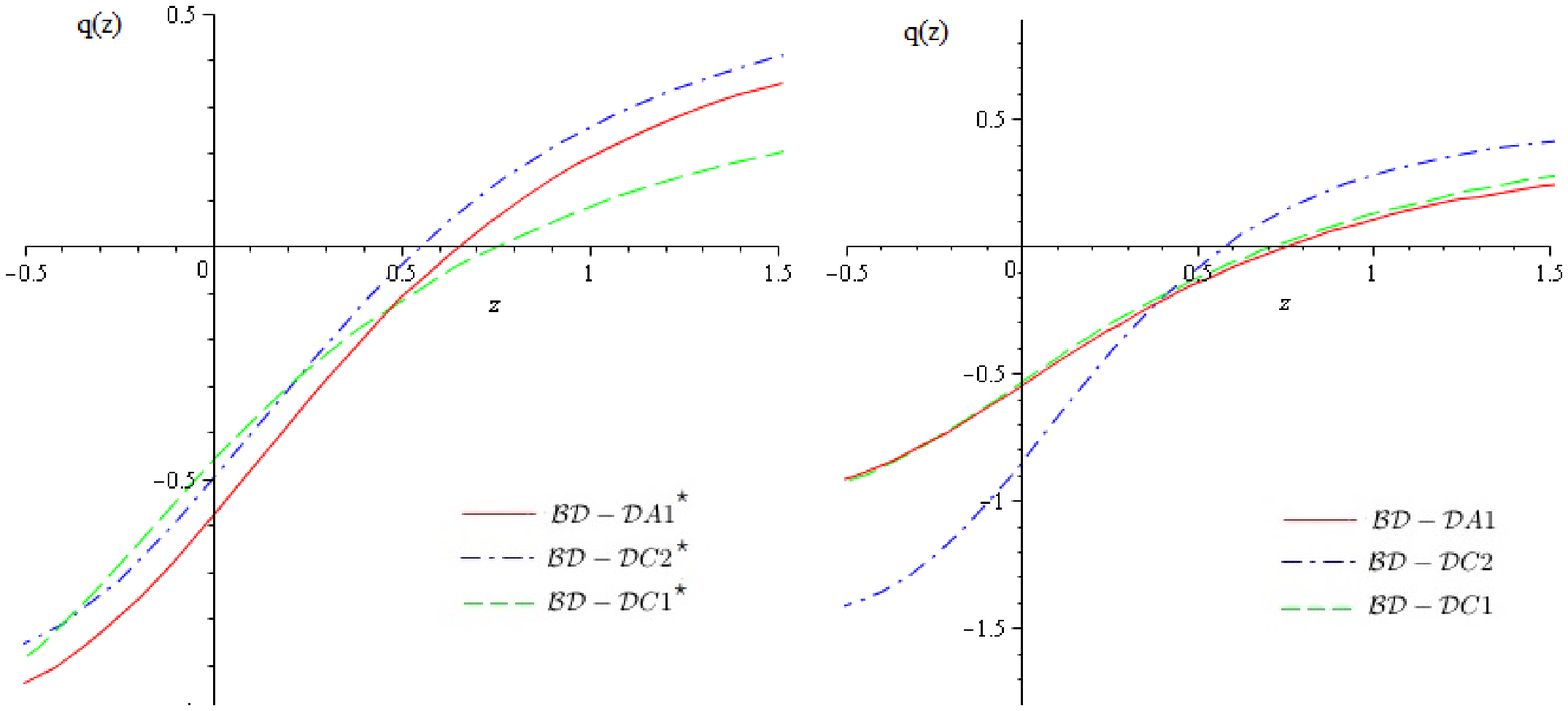}
	\caption{{\protect\footnotesize {\ Deceleration parameter, $q(z)$, versus $z$
				for interacting (right)/non-interacting (left) models according to best fitted values of Table. \ref{tableFit2}. } }}
	\label{fig:q(z)}
\end{figure*}

Let us start with first table,\ref{tableFit2}, where we have utilized BAO$_A$+SNIa+Om$h^2$ and first set of data points on $H(z_i)$. A glance at this table reveals that the fit quality
for all cases except $\mathcal{BD-D}C1^{\star }$ and $\mathcal{BD-D}C2^{\star }
$ have $\chi^2/dof$ less than $\Lambda $CDM model.
Meanwhile among all these cases, the $\mathcal{BD-D}C2$ and $\mathcal{BD-D}%
A1^{\star }$ render the best fit quality (the smallest value of $\chi
^{2}/dof$ among all others).\newline
As explained in Sec. \ref{sect:Fitting}, using $\Delta$AIC and $\Delta$BIC increments, we are
able to compare interacting and non interacting
"i"=$\mathcal{BD-D}A1$, $\mathcal{BD-D} C1$ and $\mathcal{BD-D}C2$
cases with the "j"=$\Lambda$CDM. Hence, from table \ref{tableFit2},
non-interacting $\mathcal{BD-D}A1^{\star }$ case is the only model with both positive sign for $\Delta$AIC (with very strong evidences) and $\Delta$BIC (with strong evidences) against $\Lambda$CDM. \\
Even though $\mathcal{BD-D}C2$ model shows negative sign for $\Delta$BIC, but according to $\Delta$AIC it has very strong evidences against $\Lambda$CDM. 
While for non-interacting case, $\mathcal{BD-D}C2^{\star }$, there are very strong evidences against it according to $\Delta$AIC and $\Delta$BIC which state that such model has no chance in front of $\Lambda$CDM.\\
Now, we consider $u(z)$, $w(z)$ and $q(z)$ plots for best values of Table. \ref{tableFit2}, as it is figured in Fig. \ref{fig:u(z)}, \ref{fig:w(z)} and \ref{fig:q(z)}:
The evolutionary behavior of energy density ratio is significant from the point of view of investigation of coincidence problem. As it is seen from fig. \ref{fig:u(z)} for all non-interacting models no bound is seen. While for interacting $\mathcal{BD-D}A1$ and $\mathcal{BD-D}C2$ the coincidence problem, because of finite values of $u(z)$, in past and future is alleviated which makes it as a good support for these two models.\\
In fig. \ref{fig:w(z)} the behavior of EoS parameter versus redshift is depicted. As it is transparent from this figure, all non-interacting models, except $\mathcal{BD-D}C2^{\star }$, possess an asymptotic behavior near present time. Among interacting cases, both $\mathcal{BD-D}A1$ and $\mathcal{BD-D}C2$ pass the phantom wall near present which shows more consistency with observation. Interesting prominent feature, worth noticing here to stand out in connection with EoS function is that the dynamical DE
models under study can provide a reason for the quintessence and phantom-like character of the DE without necessarily using fundamental scalar fields. So a particular interest is analysis of effective EoS of the models in this class
whose behavior near our time could explain the persistent phantom-like character of the DE
without entreating real phantom fields.\\
Another assessment we can carry out here is to compare the current value of EoS parameter for each model with what is gained by observation. By substituting the
best-fitted values of parameters from table. \ref{tableFit2} in the EoS
relation of each models current value of EoS parameter, $w_{D}^{(0)}$, has been calculated and gathered in table. \ref{tableq-z}. On the other hand,  current observational evidence of Planck 2018 results in     
$w_{D}^{(0)}=-1.028 \pm 0.032$ according to Planck TT,TE,EE+lowE+lensing+SNe+BAO, \cite{planck2018a}. A glance at table. \ref{tableq-z} shows that $\mathcal{BD-D}A1$, $\mathcal{BD-D}A1^{\star}$  and $\mathcal{BD-D}C2^{\star}$ have closest values for $w_{D}^{(0)}$ in contrast with observation.\\
The evolution of deceleration parameter, by using Eqs.
(\ref{eq:BDDA1-q}), (\ref{eq:BDDC1-q}) and (\ref{eq:BDDC2-q}), is
illustrated in Fig. \ref{fig:q(z)} for any cases. As one can see, all cases has a deflection point in the past where
the expanding universe transit from a deceleration to acceleration phase.
Deceleration parameter and deflection
 point for all $\mathcal{D}$-class cases are given in table
 \ref{tableq-z}.
This table point out that similar to the EoS parameter,
the $\mathcal{BD-D}C2$ gets the smallest value of $q^{(0)}$. Besides, transition point for $\mathcal{BD-D}A1$ model occur at farthest redshifts.

\begin{figure*}
	\centering
	\includegraphics[width=\textwidth]{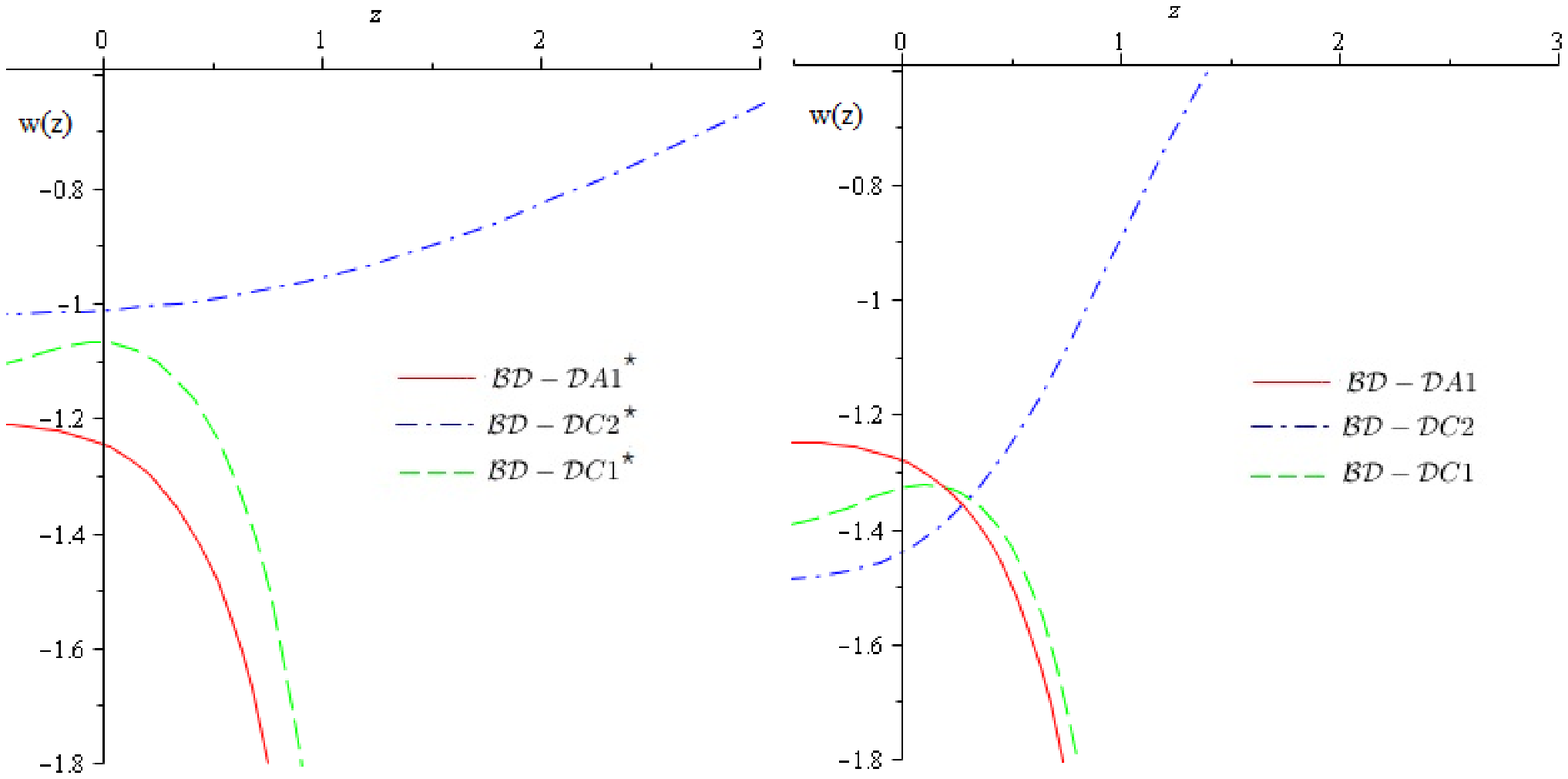}
	\caption{{\protect\footnotesize {\ The evolution of
				$\protect\omega_{D}(z)$ versus $z$, for and non-interacting (left) and interacting
				(right) models using thebest fit values of Table. \ref{tableFit3}}}} \label{fig:w(z)-2}
\end{figure*}
\begin{figure*}
	\centering
	\includegraphics[width=\textwidth]{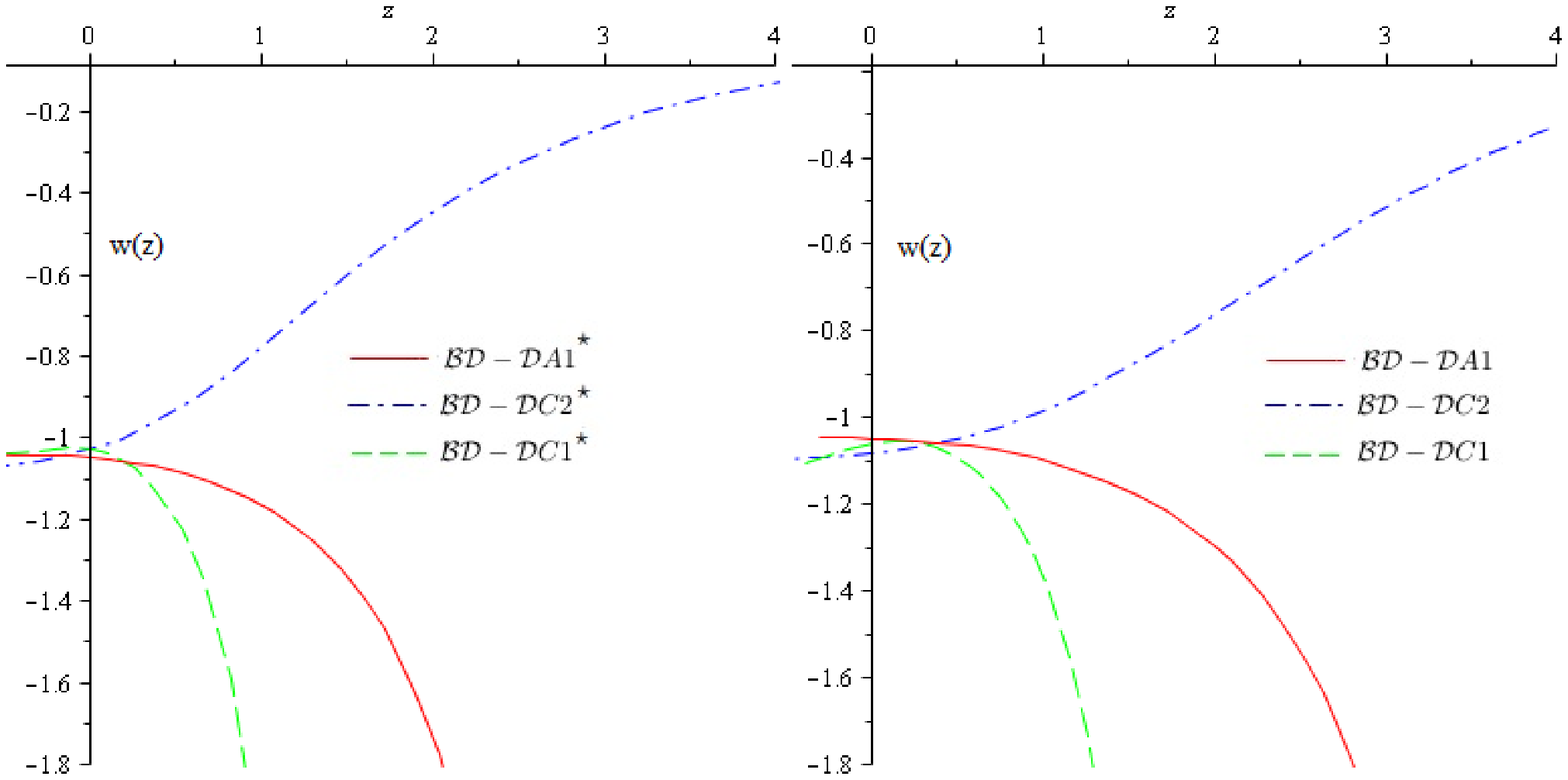}
	\caption{{\protect\footnotesize {\ The evolution of
				$\protect\omega_{D}(z)$ versus $z$, for and non-interacting (left) and interacting
				(right) models using the best fit values of Table. \ref{tableFit4}}}} \label{fig:w(z)-3}
\end{figure*}



\begin{figure*}
	\centering
	\includegraphics[totalheight=8cm]{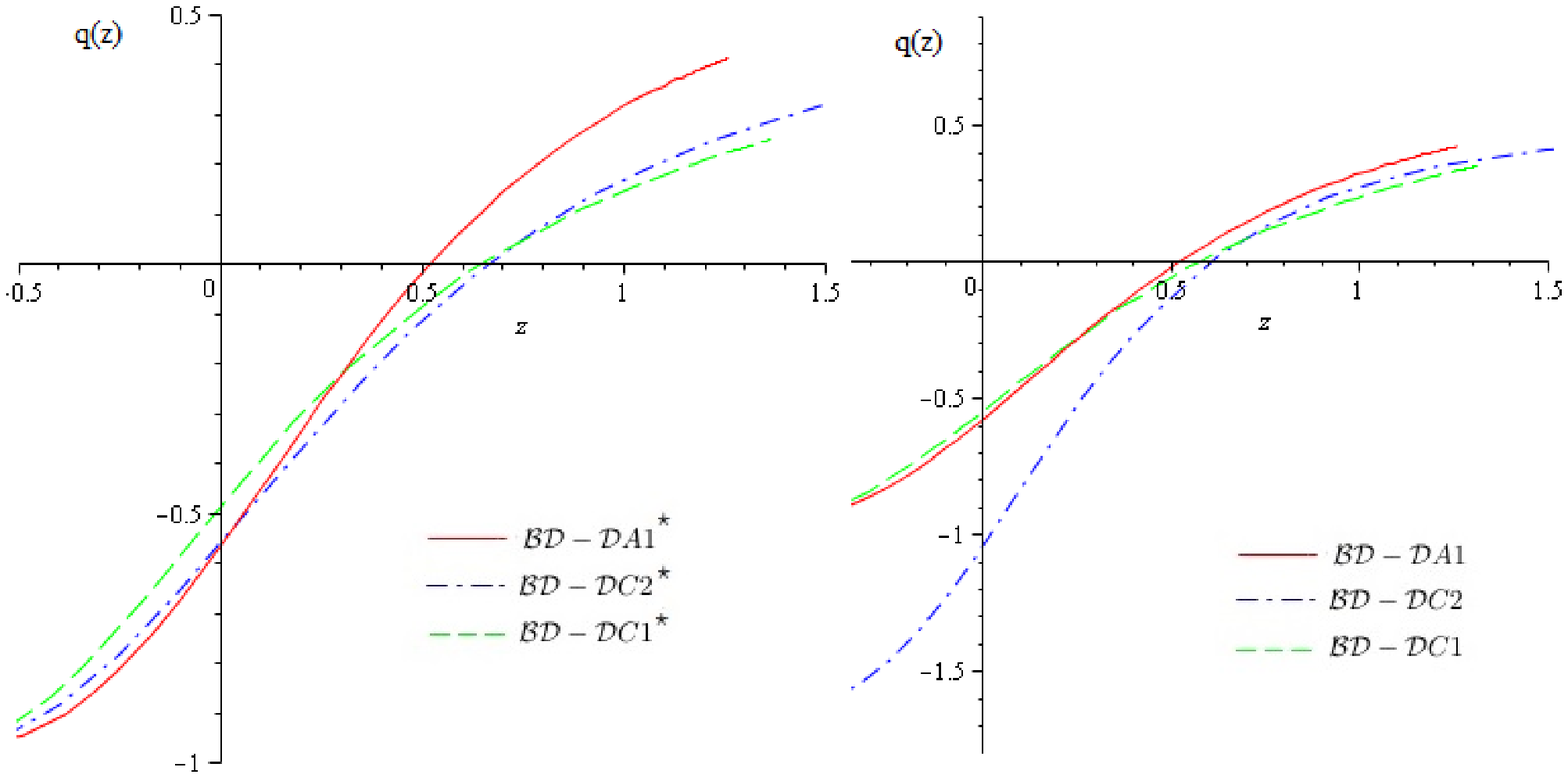}
	\caption{{\protect\footnotesize {\ Deceleration parameter, $q(z)$, versus $z$
				for interacting (right)/non-interacting (left) models according to best fit values of Table. \ref{tableFit3}}}}
	\label{fig:q(z)-2}
\end{figure*}

\begin{figure*}
	\centering
	\includegraphics[totalheight=8cm]{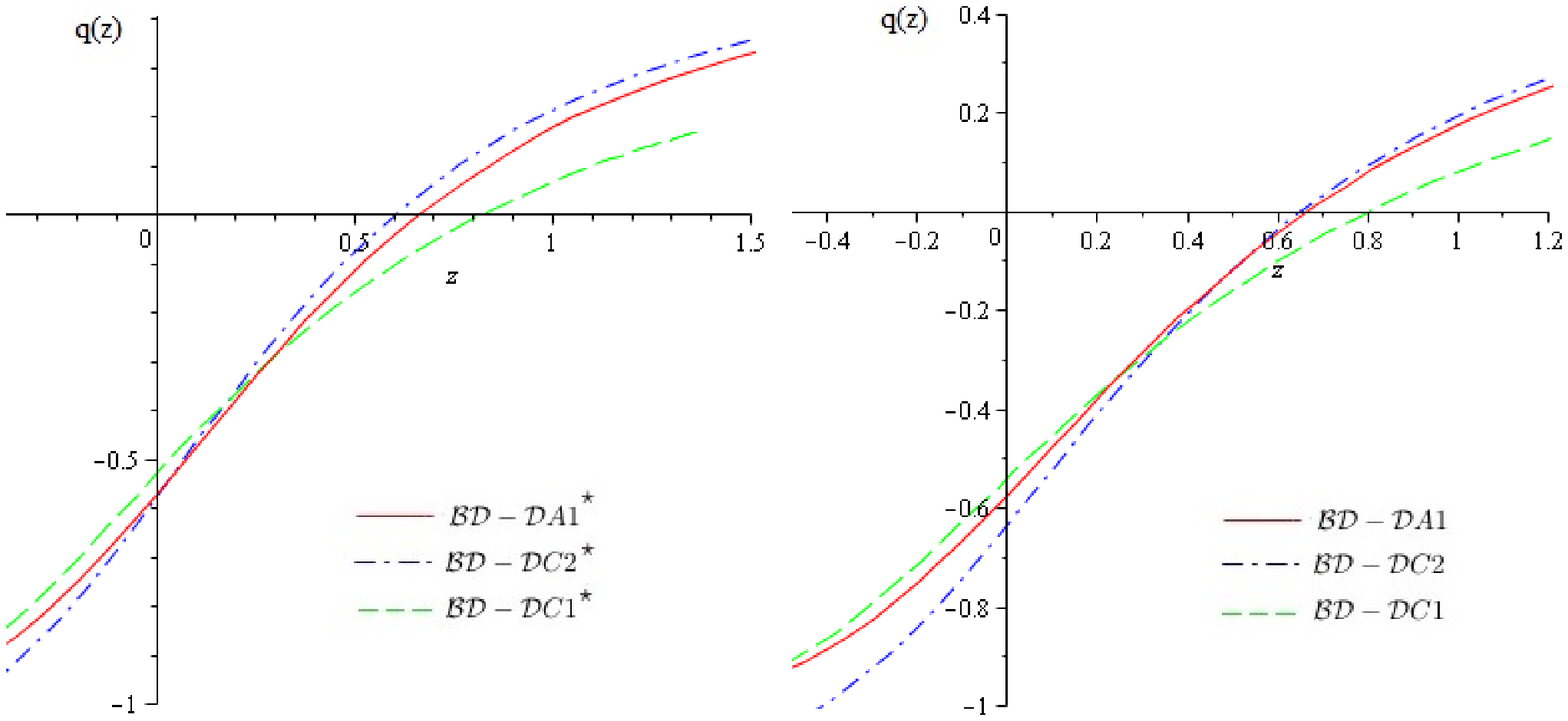}
	\caption{{\protect\footnotesize {\ Deceleration parameter, $q(z)$ versus $z$
				for interacting (right)/non-interacting (left) models according to best fit values of Table. \ref{tableFit4}}}}
	\label{fig:q(z)-3}
\end{figure*}




\begin{table*}
	\begin{center}
		\resizebox{1\textwidth}{!}{
			\begin{tabular}{| c  |c | c | c | c  | c | c  |}
				\multicolumn{1}{c}{}  & \multicolumn{1}{c}{}  & \multicolumn{1}{c}{}
				& \multicolumn{1}{c}{}  & \multicolumn{1}{c}{}  &
				\multicolumn{1}{c}{}  & \multicolumn{1}{c}{}
				\\\hline
				{}  & {\small$\mathcal{BD-D}A1$} &
				{\small$\mathcal{BD-D}A1^{\star}$} & {\small$\mathcal{BD-D}C1$
				}& {\small$\mathcal{BD-D}C1^{\star}$ } &
				{\small$\mathcal{BD-D}C2$} & {\small$\mathcal{BD-D}C2^{\star}$}
				\\\hline
				{$w^{(0)}$}  & {\small$-1.004$} & {\small$-1.086$} & {\small$-1.127$} &
				{\small$-0.909$} & {\small$-1.2960$} & {\small$-1.075$}
				\\\hline
				{$q^{(0)}$} &{\small$-0.543$} & {\small$-0.579$} & {\small$-0.529$}
				& {\small$-0.460$} & {\small$-0.844$} & {\small$-0.501$}
				\\\hline
				{$z_{tr}$}  & {\small$0.745$} & {\small$0.650$} & {\small$0.707$} &
				{\small$0.708$} & {\small$0.573$} & {\small$0.549$}
				\\\hline
			\end{tabular}}
		\end{center}
		\caption{ The present value of EoS, deceleration parameter and deflection
			point for all $\mathcal{D}$-class cases and according the best fitted values of Table. \ref{tableFit2}.} \label{tableq-z}
	\end{table*}
	
	\begin{table*}
		\begin{center}
			\resizebox{1\textwidth}{!}{
				\begin{tabular}{| c  |c | c | c | c  | c | c  |}
					\multicolumn{1}{c}{}  & \multicolumn{1}{c}{}  & \multicolumn{1}{c}{}
					& \multicolumn{1}{c}{}  & \multicolumn{1}{c}{}  &
					\multicolumn{1}{c}{}  & \multicolumn{1}{c}{}
					\\\hline
					{}  & {\small$\mathcal{BD-D}A1$} &
					{\small$\mathcal{BD-D}A1^{\star}$} & {\small$\mathcal{BD-D}C1$
					}& {\small$\mathcal{BD-D}C1^{\star}$ } &
					{\small$\mathcal{BD-D}C2$} & {\small$\mathcal{BD-D}C2^{\star}$}
					\\\hline
					{$w^{(0)}$}  & {\small$-1.280$} & {\small$-1.246$} & {\small$-1.326$} &
					{\small$-1.067$} & {\small$-1.435$} & {\small$-1.007$}
					\\\hline
					{$q^{(0)}$} &{\small$-0.577$} & {\small$-0.566$} & {\small$-0.544$}
					& {\small$-0.487$} & {\small$-1.035$} & {\small$-0.560$}
					\\\hline
					{$z_{tr}$}  & {\small$0.521$} & {\small$0.521$} & {\small$0.573$} &
					{\small$0.649$} & {\small$0.604$} & {\small$0.668$}
					\\\hline
				\end{tabular}}
			\end{center}
			\caption{ Same as in Table. \ref{tableq-z} \textbf{but} here we have used the best fitted values given from Table. \ref{tableFit3}.} \label{tableq-z-2}
		\end{table*}
		
		\begin{table*}
			\begin{center}
				\resizebox{1\textwidth}{!}{
					\begin{tabular}{| c  |c | c | c | c  | c | c  |}
						\multicolumn{1}{c}{}  & \multicolumn{1}{c}{}  & \multicolumn{1}{c}{}
						& \multicolumn{1}{c}{}  & \multicolumn{1}{c}{}  &
						\multicolumn{1}{c}{}  & \multicolumn{1}{c}{}
						\\\hline
						{}  & {\small$\mathcal{BD-D}A1$} &
						{\small$\mathcal{BD-D}A1^{\star}$} & {\small$\mathcal{BD-D}C1$
						}& {\small$\mathcal{BD-D}C1^{\star}$ } &
						{\small$\mathcal{BD-D}C2$} & {\small$\mathcal{BD-D}C2^{\star}$}
						\\\hline
						{$w^{(0)}$}  & {\small$-1.048$} & {\small$-1.048$} & {\small$-1.060$} &
						{\small$-1.028$} & {\small$-1.083$} & {\small$-1.030$}
						\\\hline
						{$q^{(0)}$} &{\small$-0.572$} & {\small$-0.572$} & {\small$-0.539$}
						& {\small$-0.530$} & {\small$-0.634$} & {\small$-0.578$}
						\\\hline
						{$z_{tr}$}  & {\small$0.665$} & {\small$0.666$} & {\small$0.800$} &
						{\small$0.821$} & {\small$0.652$} & {\small$0.601$}
						\\\hline
					\end{tabular}}
				\end{center}
				\caption{ Same as in Table. \ref{tableq-z} \textbf{but} the best fitted values are given from Table. \ref{tableFit4}.} \label{tableq-z-3}
			\end{table*}
			
			\begin{figure*}[tbp]
				\epsfxsize=17cm \centerline{\epsffile{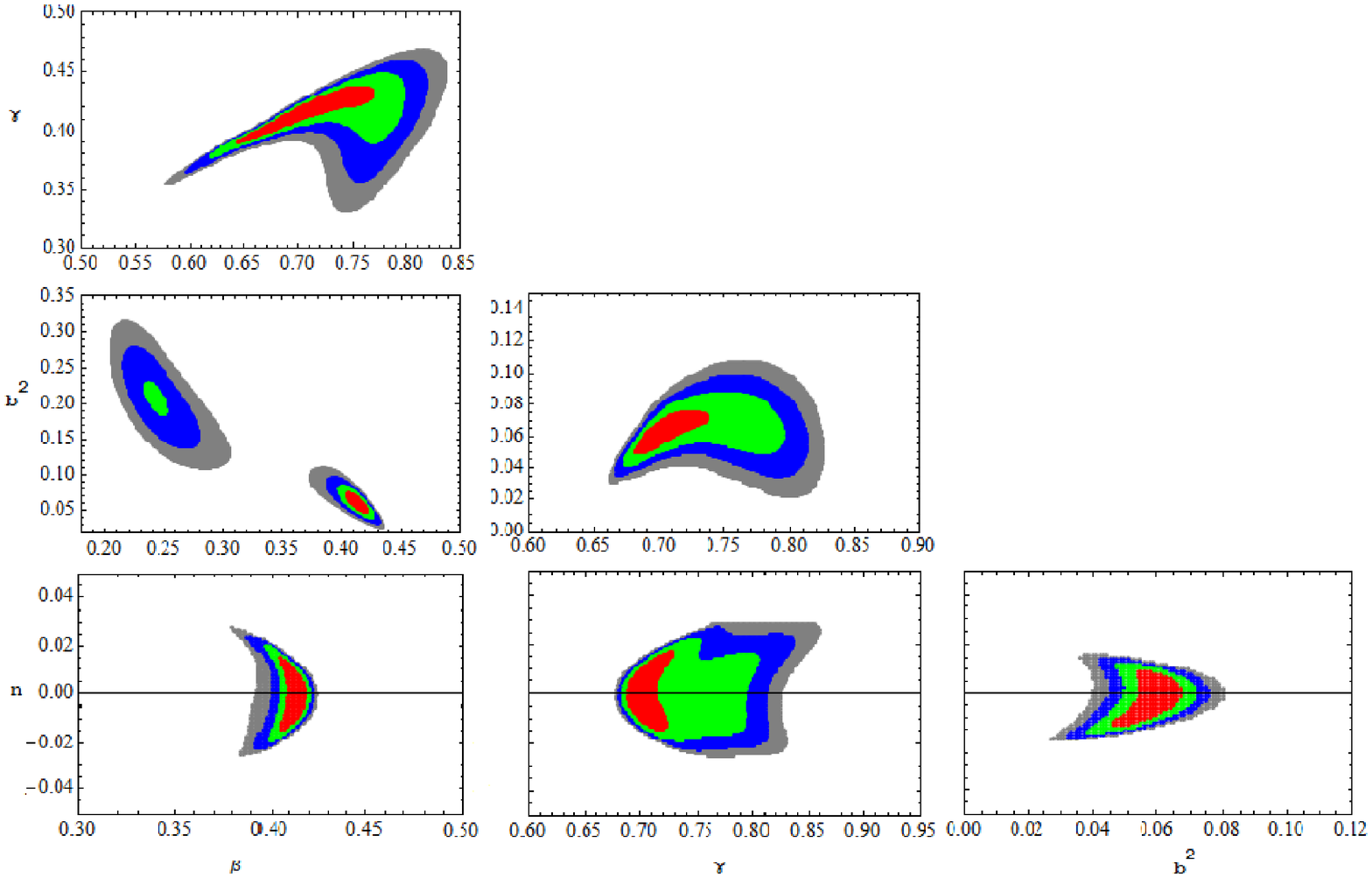}}
				\caption{{\protect\footnotesize {2-dimensional Likelihood contours of the cosmological and model parameters (for the values
							$-2\ln\mathcal{L}/\mathcal{ L}_{max}=2.30$, $6.16, 11.81, 19.33$,	corresponding to 1$\protect\sigma$, 2$\protect \sigma$,
							$3\protect\sigma$ and $4\protect\sigma$ confidence levels) for the $\mathcal{BD-D}C2$
							model using the full expansion history: (Omh$^2$+$BAO_A$+SNIa) and with second set of data.
						} }} \label{fig:contour plot)}
					\end{figure*}

Now let investigate the results according to table \ref{tableFit3}. For obtaining the best fitted values of this table, we have applied the expansion history: BAO$_A$+SNIa+Om$h^2$ and with second set of data points on $H(z_i)$  obtained by differential-age techniques which has no correlation with BAO data. We have applied both set of data for BAO$_A$+SNIa+Om$h^2$ analysis to better see the effect of data on final result of best fit values.\\ Regarding this table, both $\Delta$AIC and $\Delta$BIC shows "very strong evidences" for $\mathcal{BD-D}A1$ and $\mathcal{BD-D}A1^{\star}$. Whereas, we see " very strong evidences" against $\mathcal{BD-D}C1^{\star}$  and $\mathcal{BD-D}C1$ in front of $\Lambda CDM$.  While for $\mathcal{BD-D}C2$ there is "strong evidences" according to $\Delta$AIC .\\
Now for checking the background effects of this data analyze, we take a look in Figs. \ref{fig:u(z)-2}, \ref{fig:w(z)-2} and \ref{fig:q(z)-2}, which is gained by the best fitted values of table. \ref{tableFit3}.
The $u(z)$ plot shows alleviation of coincidence problem for both interacting and non-interacting $\mathcal{BD-D}C2$ models. Also passing phantom wall will occur just for these two models according to fig.\ref{fig:w(z)-2}. However,  $w_{D}^{(0)}$ for $\mathcal{BD-D}C1^{\star}$ and 
$\mathcal{BD-D}C2^{\star}$ is close to observational value of $w_{D}^{(0)}$, but as it is seen $\mathcal{BD-D}C1^{\star}$ very strongly dis-proofed by both increments . On the other hand, from table \ref{tableq-z-2}, $\mathcal{BD-D}C2$ model has smallest value for deceleration parameter at present. \\

Finally, lets discuss about the result in table. \ref{tableFit4} where we have used the expansion history: BAO$_A$+SNIa+H(z) and with second set of data points on $H(z_i)$. Surprisingly this table shows negative sign for all models and for both increments. Just we may provide some clues for $\mathcal{BD-D}A1^{\star}$ against $ \Lambda$CDM as it has the AIC value less than 2 which says no significant evidence for or against this model.
So we behold that such data and with $H(z)$ analyses instead of $omh^2$ for $H(z_i)$  could not discriminate very well between models.\\
Considering background plots, which is depicted by use of the best fitted values of this table. \ref{tableFit4}, we perceive that for u(z) both interacting and non-interacting $\mathcal{BD-D}C2$ have finite values in past and future and smooth the coincidence problem. Even though AIC and BIC are against these models but the positive point here is that the obtained best fitted values according to this analysis and with second set of data on H(z) causes to have reduction for u(z) for interacting and non-interacting $\mathcal{BD-D}C2$ cases. Besides,  fig. \ref{fig:w(z)-3} indicate again here that just for these two models, $\mathcal{BD-D}C2$ and $\mathcal{BD-D}C2^{\star}$, the EoS parameter cross phantom wall and has $w(z)\geq-1$ in past. \\
 While the current values of EoS parameter according to table. \ref{tableq-z-3}, are close to observation but among all these models, $\mathcal{BD-D}C1^{\star}$ and  $\mathcal{BD-D}C2^{\star}$ posses the most closeness  and are perfectly compatible with observation, ref. \cite{planck2018a}.\\ 
Finally, plot. \ref{fig:q(z)-3} shows two $\mathcal{BD-D}C2$ and $\mathcal{BD-D}C2^{\star}$ models have less values of deceleration parameter at current time and also less $z_{tr}$ values which all these also could be checked by table. \ref{tableq-z-3}.  \\ 
We gather the results of Table. \ref{tableFit4} by mentioning that even though both increments have no positive signs for these $\mathcal{BD-D}$ models, but the background investigation reveals soothed behavior of u(z) for interacting and non-interacting $\mathcal{BD-D}C2$. 
Another salient property which evinced during the composition of BAO$_A$+SNIa+H(z) analyze with second set of data points on $H(z_i)$ is that free parameter, n , for all models grabs positive sign. While one can check that for two other tables parameter "n" has positive or negative sign depend the models.\\

Now we concentrate on Fig. \ref{fig:contour plot)}, where the
2-dimensional plots for the physical region of parameters of
$\mathcal{BD-D}C2$ has been demonstrated. We have utilized the expansion history data (Omh$^{2}$+$BAO_{A}$+SNIa) and second set of data points for Hubble parameters. The bounds with elliptically shapes corresponds with $1\sigma $, $%
2\sigma $, $3\sigma $ and $4\sigma $ confidence level. \\


To close this section, we collect consequences of all tables for each model. But before,  it is important to
 underline here that using both sets of data points and also two different diagnostic, i.e. $omh^2$ and H(z), enable a particular feature to compare analytically the gained results associated with each analyze and each data sets. Besides it illustrates which diagnostics could better determine the consistency of each models and better discriminate between all introduced Hubble-rate-dependent dynamical DE
 cases in this paper. In the following we sum the results up for all models separately  and by remarking all three analyzes and both sets of data for Hubble parameter: \\
 
1- $\mathcal{BD-D}A1$ model (both interacting and non-interacting):\\
Non-interacting case according to BAO$_A$+SNIa+$omh^2$ and both set of  data points on H(z), tables. \ref{tableFit2} and \ref{tableFit3}, has very strong evidences against  $\Lambda CDM$. Also its the only model that, by assuming BAO$_A$+SNIa+H(z) analyze and with first set of data, displays some hopes against $\Lambda CDM$ (as other models have all negative big sign in front of $\Lambda CDM$).
So $\mathcal{BD-D}A1^{\star}$ is most promising model among all according to results of our statistical analysis here in this paper. \\
The interacting one exhibits strong evidences against $\Lambda CDM$ in BAO$_A$+SNIa+$omh^2$ and with second set of data and as the same way evidences ($\geq2$)  against  $\Lambda CDM$ regarding the same analyze but with first set of data on H(z). Good to emphasize here that u(z) just for interacting $\mathcal{BD-D}A1$ and pursuant to table. \ref{tableFit2} shows alleviation for coincidence problem.\\
Reviewing plots for w(z), Figs. \ref{fig:w(z)}, \ref{fig:w(z)-2}, \ref{fig:w(z)-3}, one can see that interacting $\mathcal{BD-D}A1$ with best fitted values of first table has this ability to cross from quintessence to phantom in past. Also its current value of EoS has most consistency with observation.\\
Hereupon interacting $\mathcal{BD-D}A1$ with first analyze and 
first set of data on Hubble parameter has this merit that simultaneously  shows better analyze, moderates the coincidence problem and at the same time its EoS parameter presents very good consistency with observation.\\
Eventually, it is good to keep in mind though that non-interacting $\mathcal{BD-D}A1^{\star}$  renders a perfect analyze with very strong evidences against $\Lambda CDM$ even though the coincidence problem remain valid yet in non-interacting case.\\

2- $\mathcal{BD-D}C1$ model (both interacting and non-interacting):\\ 
Except $\Delta$AIC for interacting $\mathcal{BD-D}C1$ in first table. \ref{tableFit2}, the sign for both AIC and BIC and for all three tables are negative. 
Outcomes from background history investigations and plots for EoS and u(z) parameters reveal phenomenologically problematic issues (coincidence problem and inconsistency with current observational data) for this model.
Due to these obstacles, this model does not possess the ability for
proper adjustment with expansion history of universe and could be
ruled out. Furthermore, it is interesting to know that such model has been ruled out in the context of general relativity according to ref. \citep*{Gomez-Valent:2015pia}.\\

3- $\mathcal{BD-D}C2$ model (both interacting and non-interacting):\\ 
Non-interacting case: confronting with BIC in all three tables, \ref{tableFit2}, \ref{tableFit3} and \ref{tableFit4}, there are very strong, strong and very strong evidences  in turn against this model. AIC increment is against this model too but just regarding table \ref{tableFit3} it shows the value of $0.717$ which says it has no cons or pros in comparison with $\Lambda CDM$.
Advantages of $\mathcal{BD-D}C2^{\star}$ are alleviation of coincidence problem according to u(z) plots, \ref{fig:u(z)-2} and \ref{fig:u(z)-3} (using the best fitted values of tables. \ref{tableFit3} and \ref{tableFit4}). Also it depicts cross from quintessence to phantom  for all best fitted values and besides its current amount of EoS is not so far from observation. By the way as it is discussed before both increments manifest that it does not present statistically adequate result versus $\Lambda$CDM.\\
interacting case: it indicates strong and very strong evidences against $\Lambda CDM$ and regarding tables. \ref{tableFit2} and \ref{tableFit3} and its AIC. But BIC is against this model in light of all analyzes in this paper.
Then again, interacting $\mathcal{BD-D}C2$ soften the coincidence problem as well for all three analyzes and both sets af data on H(z).  It exposes quintessence to phantom cross for all best fitted values and mimics observational results for background history  very well.\\


At the end of this section, let us conclude by emphasizing our main message.
In view of tables. \ref{tableFit2} and \ref{tableFit3}, $\mathcal{BD-D}A1^{\star}$ model get the best position according to both $AIC$ and $BIC$ increments with very strong evidences. Then the second position is grabbed by $\mathcal{BD-D}C2$ model from the point of view of $AIC$ which shows strong evidence against $\Lambda CDM$. But then, between these two models and in light of plots for $u(z)$ just $\mathcal{BD-D}C2$ could diminish the coincidence problem. Besides, plots for EoS indicate that only this model has ability to pass from quintessence to phantom regarding all three statistical significance tables. \ref{tableFit2}, \ref{tableFit3} and \ref{tableFit4}.\\
Definitely, Structure formation analysis could better distinguished between these two models which will be the subject of future works.
But what is apparent form our analyzes here and without need to structure formation study is that both interacting and non-interacting $\mathcal{BD-D}C1$ models have large incongruity with both background history analysis and both data sets and  must be abandoned indispensably.


\section{Conclusions}
Three cases of $\mathcal{D}$-class, interacting and
non-interacting, of dark energy investigated in the context of
Brans-Dike theory of gravity. The Hubble rate, equation of
state and deceleration parameters are given and showed that the
cosmic coincidence problem may be alleviated in some cases and almost in interacting ones. \\
In this paper, we have exerted both diagnostics for Hubble parameter i.e.
 $Omh^{2}$ and $H(z)$ via SNIa+BAO$_A$+$Omh^{2}$ and SNIa+BAO$_A$+$H(z)$
 analysis to could remark their effects and discrepancy on final results.
 Also in order to present the effects of data points in each 
diagnostics we have utilized correlated and uncorrelated data sets of H(z)
 parameter in SNIa+BAO$_A$+$Omh^{2}$ fit and just uncorrelated one in
SNIa+BAO$_A$+$H(z)$ analyze. 
The outcomes entirely presented in tables and
have been compared with $\Lambda$CDM model. So after a detailed study we
found following facts:\\

Non-interacting $\mathcal{BD-D}A1^{\star}$ subclass exhibits striking statistical analysis among all other models and against $\Lambda$CDM. 
While interacting $\mathcal{BD-D}A1$ subclass, utilizing SNIa+BAO$_A$+$Omh^{2}$ and second set of data set for H(z), expose admissible statistical analyze but just by considering SNIa+BAO$_A$+$Omh^{2}$ and first set of data set for H(z) this subclass acquire the ability to pass phantom wall and mitigate the coincidence problem.\\

Notable result for both interacting and noninteracting $\mathcal{BD-D}C2$ models is that such model have capability to mimic the quintessence
behavior of EoS and provide a possible explanation for the phantom character of the DE
at present for both data sets and all three analysis.
Besides, interacting one has also this potency to alleviate coincidence problem in all cases and according to all analyses.\\
 
Using the same testing tools we have reached the firm conclusion that both interacting and non-interacting $\mathcal{BD-D}C1$  models are strongly disfavored and become automatically excluded by our analysis. 
Significant result which is apparent from our fit and without need to structure formation analysis is that the $\mathcal{BD-D}C1$ is not consistent with cosmic background and must be ruled out. 
 .\\ 
 
At the end of the day the most distinguished dynamical $\mathcal{BD-D}$-models, both theoretically and
phenomenologically, are those in the $\mathcal{BD-D}A1$ and $\mathcal{BD-D}C2$ classes. The fit quality rendered by them has been shown to be significantly better than that
of the $\Lambda$CDM.  These models improve significantly the fit quality of the $\Lambda$CDM, presenting that a smooth dynamical DE
behavior is better than having a rigid $\Lambda$-term for the overall cosmic history.\\
While $\mathcal{BD-D}A1$ is acceptable from our statistical point of view,  on the other hand $\mathcal{BD-D}C2$ model is considerable for alleviation of coincidence problem and good mimic of background history.
It exhibits somehow competent analyze via AIC but not as qualified as $\mathcal{BD-D}A1$ model.\\
So structure formation analysis may finally distinguished between these models in better way and we expect that the outcomes achieved here also be confirmed after
studying on the structure formation analysis of these models. We leave this for future works.\\

\acknowledgments

 We would like to express sincere gratitude to Joan Sol\`{a} for constructive
 comments and discussion. E. Karimkhani would also like to
 thank  Adri\`a G\'omez-Valent for sharing his knowledge on data fitting procedure.

\bibliographystyle{spr-mp-nameyear-cnd}

\bibliography{refbd}

\begin{thebibliography}{51}
\ifx \bisbn   \undefined \def \bisbn  #1{ISBN #1}\fi
\ifx \binits  \undefined \def \binits#1{#1} \fi
\ifx \bauthor  \undefined \def \bauthor#1{#1} \fi
\ifx \batitle  \undefined \def \batitle#1{#1} \fi
\ifx \bjtitle  \undefined \def \bjtitle#1{#1}\fi
\ifx \bvolume  \undefined \def \bvolume#1{\textbf{#1}}\fi
\ifx \byear  \undefined \def \byear#1{#1} \fi
\ifx \bissue  \undefined \def \bissue#1{#1} \fi
\ifx \bfpage  \undefined \def \bfpage#1{#1} \fi
\ifx \blpage  \undefined \def \blpage #1{#1} \fi
\ifx \burl  \undefined \def \burl#1{\textsf{#1}} \fi
\ifx \doiurl  \undefined \def \doiurl#1{\textsf{#1}} \fi
\ifx \betal  \undefined \def \betal{\textit{et al.}} \fi
\ifx \binstitute  \undefined \def \binstitute#1{#1} \fi
\ifx \binstitutionaled  \undefined \def \binstitutionaled#1{#1} \fi
\ifx \bctitle  \undefined \def \bctitle#1{#1} \fi
\ifx \beditor  \undefined \def \beditor#1{#1} \fi
\ifx \bpublisher  \undefined \def \bpublisher#1{#1} \fi
\ifx \bbtitle  \undefined \def \bbtitle#1{#1} \fi
\ifx \bedition  \undefined \def \bedition#1{#1} \fi
\ifx \bseriesno  \undefined \def \bseriesno#1{#1} \fi
\ifx \blocation  \undefined \def \blocation#1{#1} \fi
\ifx \bsertitle  \undefined \def \bsertitle#1{#1} \fi
\ifx \bsnm \undefined \def \bsnm#1{#1} \fi
\ifx \bsuffix \undefined \def \bsuffix#1{#1} \fi
\ifx \bparticle \undefined \def \bparticle#1{#1} \fi
\ifx \barticle \undefined \def \barticle#1{#1} \fi
\ifx \bconfdate \undefined \def \bconfdate #1{#1} \fi
\ifx \botherref \undefined \def \botherref #1{#1} \fi
\ifx \url \undefined \def \url#1{\textsf{#1}} \fi
\ifx \bchapter \undefined \def \bchapter#1{#1} \fi
\ifx \bbook \undefined \def \bbook#1{#1} \fi
\ifx \bcomment \undefined \def \bcomment#1{#1} \fi
\ifx \oauthor \undefined \def \oauthor#1{#1} \fi
\ifx \citeauthoryear \undefined \def \citeauthoryear#1{#1} \fi
\ifx \endbibitem  \undefined \def \endbibitem {}\fi
\ifx \bconflocation  \undefined \def \bconflocation#1{#1} \fi
\ifx \arxivurl  \undefined \def \arxivurl#1{\textsf{#1}} \fi

\bibitem[\protect\citeauthoryear{Acquaviva and Verde}{2007}]{Acquaviva:2007mm}
\begin{barticle}
\bauthor{\bsnm{Acquaviva}, \binits{V.}},
\bauthor{\bsnm{Verde}, \binits{L.}}:
\bjtitle{JCAP}
\bvolume{0712},
\bfpage{001}
(\byear{2007}).
\arxivurl{arXiv:0709.0082}.
doi:\doiurl{10.1088/1475-7516/2007/12/001}
\end{barticle}
\endbibitem

\bibitem[\protect\citeauthoryear{Ade et~al.}{2016}]{Ade:2015xua}
\begin{barticle}
\bauthor{\bsnm{Ade}, \binits{P.A.R.}}, \betal:
\bjtitle{Astron. Astrophys.}
\bvolume{594},
\bfpage{13}
(\byear{2016}).
\arxivurl{arXiv:1502.01589}.
doi:\doiurl{10.1051/0004-6361/201525830}
\end{barticle}
\endbibitem

\bibitem[\protect\citeauthoryear{Aghanim et~al.}{}]{planck2018a}
\begin{botherref}
\oauthor{\bsnm{Aghanim}, \binits{N.}}, et al.:
{Planck2018 results. VI. Cosmological parameters}.
\arxivurl{arXiv:1807.06209}
\end{botherref}
\endbibitem

\bibitem[\protect\citeauthoryear{{Akaike}}{1974}]{1974ITAC...19..716A}
\begin{barticle}
\bauthor{\bsnm{{Akaike}}, \binits{H.}}:
\bjtitle{IEEE Transactions on Automatic Control}
\bvolume{19},
\bfpage{716}
(\byear{1974})
\end{barticle}
\endbibitem

\bibitem[\protect\citeauthoryear{Alavirad and Sheykhi}{2014}]{Alavirad:2014kqa}
\begin{barticle}
\bauthor{\bsnm{Alavirad}, \binits{H.}},
\bauthor{\bsnm{Sheykhi}, \binits{A.}}:
\bjtitle{Phys. Lett.}
\bvolume{B734},
\bfpage{148}
(\byear{2014}).
\arxivurl{arXiv:1405.2515}.
doi:\doiurl{10.1016/j.physletb.2014.05.023}
\end{barticle}
\endbibitem

\bibitem[\protect\citeauthoryear{{Arnett}}{1982a}]{1982ApJ...254....1A}
\begin{barticle}
\bauthor{\bsnm{{Arnett}}, \binits{W.D.}}:
\bjtitle{\apj}
\bvolume{254},
\bfpage{1}
(\byear{1982}a).
doi:\doiurl{10.1086/159698}
\end{barticle}
\endbibitem

\bibitem[\protect\citeauthoryear{{Arnett}}{1982b}]{1982ApJ...253..785A}
\begin{barticle}
\bauthor{\bsnm{{Arnett}}, \binits{W.D.}}:
\bjtitle{\apj}
\bvolume{253},
\bfpage{785}
(\byear{1982}b).
doi:\doiurl{10.1086/159681}
\end{barticle}
\endbibitem

\bibitem[\protect\citeauthoryear{Avilez and Skordis}{2014}]{Avilez:2013dxa}
\begin{barticle}
\bauthor{\bsnm{Avilez}, \binits{A.}},
\bauthor{\bsnm{Skordis}, \binits{C.}}:
\bjtitle{Phys. Rev. Lett.}
\bvolume{113}(\bissue{1}),
\bfpage{011101}
(\byear{2014}).
\arxivurl{arXiv:1303.4330}.
doi:\doiurl{10.1103/PhysRevLett.113.011101}
\end{barticle}
\endbibitem

\bibitem[\protect\citeauthoryear{Banerjee and Pavon}{2001}]{Banerjee:2000gt}
\begin{barticle}
\bauthor{\bsnm{Banerjee}, \binits{N.}},
\bauthor{\bsnm{Pavon}, \binits{D.}}:
\bjtitle{Class. Quant. Grav.}
\bvolume{18},
\bfpage{593}
(\byear{2001}).
\arxivurl{arXiv:gr-qc/0012098}.
doi:\doiurl{10.1088/0264-9381/18/4/302}
\end{barticle}
\endbibitem

\bibitem[\protect\citeauthoryear{Basilakos et~al.}{2009}]{Basilakos:2009wi}
\begin{barticle}
\bauthor{\bsnm{Basilakos}, \binits{S.}},
\bauthor{\bsnm{Plionis}, \binits{M.}},
\bauthor{\bsnm{Sol\`{a}}, \binits{J.}}:
\bjtitle{Phys. Rev.}
\bvolume{D80},
\bfpage{083511}
(\byear{2009}).
\arxivurl{arXiv:0907.4555}.
doi:\doiurl{10.1103/PhysRevD.80.083511}
\end{barticle}
\endbibitem

\bibitem[\protect\citeauthoryear{Bertotti et~al.}{2003}]{Bertotti:2003rm}
\begin{barticle}
\bauthor{\bsnm{Bertotti}, \binits{B.}},
\bauthor{\bsnm{Iess}, \binits{L.}},
\bauthor{\bsnm{Tortora}, \binits{P.}}:
\bjtitle{Nature}
\bvolume{425},
\bfpage{374}
(\byear{2003}).
doi:\doiurl{10.1038/nature01997}
\end{barticle}
\endbibitem

\bibitem[\protect\citeauthoryear{Blake et~al.}{2011}]{Blake:2011en}
\begin{barticle}
\bauthor{\bsnm{Blake}, \binits{C.}}, \betal:
\bjtitle{Mon. Not. Roy. Astron. Soc.}
\bvolume{418},
\bfpage{1707}
(\byear{2011}).
\arxivurl{arXiv:1108.2635}.
doi:\doiurl{10.1111/j.1365-2966.2011.19592.x}
\end{barticle}
\endbibitem

\bibitem[\protect\citeauthoryear{Branch}{2001}]{Branch:2000zm}
\begin{barticle}
\bauthor{\bsnm{Branch}, \binits{D.}}:
\bjtitle{AIP Conf. Proc.}
\bvolume{565}(\bissue{1}),
\bfpage{31}
(\byear{2001}).
\arxivurl{arXiv:astro-ph/0012300}.
doi:\doiurl{10.1063/1.1377070}
\end{barticle}
\endbibitem

\bibitem[\protect\citeauthoryear{Brans and Dicke}{1961}]{Brans:1961sx}
\begin{barticle}
\bauthor{\bsnm{Brans}, \binits{C.}},
\bauthor{\bsnm{Dicke}, \binits{R.H.}}:
\bjtitle{Phys. Rev.}
\bvolume{124},
\bfpage{925}
(\byear{1961}).
doi:\doiurl{10.1103/PhysRev.124.925}
\end{barticle}
\endbibitem

\bibitem[\protect\citeauthoryear{Chen and Kamionkowski}{1999}]{Chen:1999qh}
\begin{barticle}
\bauthor{\bsnm{Chen}, \binits{X.-l.}},
\bauthor{\bsnm{Kamionkowski}, \binits{M.}}:
\bjtitle{Phys. Rev.}
\bvolume{D60},
\bfpage{104036}
(\byear{1999}).
\arxivurl{arXiv:astro-ph/9905368}.
doi:\doiurl{10.1103/PhysRevD.60.104036}
\end{barticle}
\endbibitem

\bibitem[\protect\citeauthoryear{Copeland et~al.}{2006}]{Copeland:2006wr}
\begin{barticle}
\bauthor{\bsnm{Copeland}, \binits{E.J.}},
\bauthor{\bsnm{Sami}, \binits{M.}},
\bauthor{\bsnm{Tsujikawa}, \binits{S.}}:
\bjtitle{Int. J. Mod. Phys.}
\bvolume{D15},
\bfpage{1753}
(\byear{2006}).
\arxivurl{arXiv:hep-th/0603057}.
doi:\doiurl{10.1142/S021827180600942X}
\end{barticle}
\endbibitem

\bibitem[\protect\citeauthoryear{de~Cruz~Perez and
  Sol\`{a}~Peracaula}{2018}]{Javier:2018sol}
\begin{barticle}
\bauthor{\bparticle{de} \bsnm{Cruz~Perez}, \binits{J.}},
\bauthor{\bsnm{Sol\`{a}~Peracaula}, \binits{J.}}:
\bjtitle{Modern Physics Letters A}
\bvolume{33},
\bfpage{1850228}
(\byear{2018}).
\arxivurl{arXiv:1809.03329}.
doi:\doiurl{10.1142/S0217732318502280}
\end{barticle}
\endbibitem

\bibitem[\protect\citeauthoryear{Delubac et~al.}{2015}]{Delubac:2014aqe}
\begin{barticle}
\bauthor{\bsnm{Delubac}, \binits{T.}}, \betal:
\bjtitle{Astron. Astrophys.}
\bvolume{574},
\bfpage{59}
(\byear{2015}).
\arxivurl{arXiv:1404.1801}.
doi:\doiurl{10.1051/0004-6361/201423969}
\end{barticle}
\endbibitem

\bibitem[\protect\citeauthoryear{Dicke}{1962}]{Dicke:1961gz}
\begin{barticle}
\bauthor{\bsnm{Dicke}, \binits{R.H.}}:
\bjtitle{Phys. Rev.}
\bvolume{125},
\bfpage{2163}
(\byear{1962}).
doi:\doiurl{10.1103/PhysRev.125.2163}
\end{barticle}
\endbibitem

\bibitem[\protect\citeauthoryear{Ding et~al.}{2015}]{Ding:2015vpa}
\begin{barticle}
\bauthor{\bsnm{Ding}, \binits{X.}},
\bauthor{\bsnm{Biesiada}, \binits{M.}},
\bauthor{\bsnm{Cao}, \binits{S.}},
\bauthor{\bsnm{Li}, \binits{Z.}},
\bauthor{\bsnm{Zhu}, \binits{Z.-H.}}:
\bjtitle{Astrophys. J.}
\bvolume{803}(\bissue{2}),
\bfpage{22}
(\byear{2015}).
\arxivurl{arXiv:1503.04923}.
doi:\doiurl{10.1088/2041-8205/803/2/L22}
\end{barticle}
\endbibitem

\bibitem[\protect\citeauthoryear{Eisenstein et~al.}{2005}]{Eisenstein:2005su}
\begin{barticle}
\bauthor{\bsnm{Eisenstein}, \binits{D.J.}}, \betal:
\bjtitle{Astrophys. J.}
\bvolume{633},
\bfpage{560}
(\byear{2005}).
\arxivurl{arXiv:astro-ph/0501171}.
doi:\doiurl{10.1086/466512}
\end{barticle}
\endbibitem

\bibitem[\protect\citeauthoryear{Farooq and Ratra}{2013}]{Farooq:2013hq}
\begin{barticle}
\bauthor{\bsnm{Farooq}, \binits{O.}},
\bauthor{\bsnm{Ratra}, \binits{B.}}:
\bjtitle{Astrophys. J.}
\bvolume{766},
\bfpage{7}
(\byear{2013}).
\arxivurl{arXiv:1301.5243}.
doi:\doiurl{10.1088/2041-8205/766/1/L7}
\end{barticle}
\endbibitem

\bibitem[\protect\citeauthoryear{Gomez-Gomar et~al.}{1998}]{GomezGomar:1997iv}
\begin{barticle}
\bauthor{\bsnm{Gomez-Gomar}, \binits{J.}},
\bauthor{\bsnm{Isern}, \binits{J.}},
\bauthor{\bsnm{Jean}, \binits{P.}}:
\bjtitle{Mon. Not. Roy. Astron. Soc.}
\bvolume{295},
\bfpage{1}
(\byear{1998}).
\arxivurl{arXiv:astro-ph/9709048}.
doi:\doiurl{10.1046/j.1365-8711.1998.29511115.x}
\end{barticle}
\endbibitem

\bibitem[\protect\citeauthoryear{G\'omez-Valent and
  Sol\`{a}}{2015}]{Gomez-Valent:2014fda}
\begin{barticle}
\bauthor{\bsnm{G\'omez-Valent}, \binits{A.}},
\bauthor{\bsnm{Sol\`{a}}, \binits{J.}}:
\bjtitle{Mon. Not. Roy. Astron. Soc.}
\bvolume{448},
\bfpage{2810}
(\byear{2015}).
\arxivurl{arXiv:1412.3785}.
doi:\doiurl{10.1093/mnras/stv209}
\end{barticle}
\endbibitem

\bibitem[\protect\citeauthoryear{G\'omez-Valent
  et~al.}{2015a}]{Gomez-Valent:2015pia}
\begin{barticle}
\bauthor{\bsnm{G\'omez-Valent}, \binits{A.}},
\bauthor{\bsnm{Karimkhani}, \binits{E.}},
\bauthor{\bsnm{Sol\`{a}}, \binits{J.}}:
\bjtitle{JCAP}
\bvolume{1512}(\bissue{12}),
\bfpage{048}
(\byear{2015}a).
\arxivurl{arXiv:1509.03298}.
doi:\doiurl{10.1088/1475-7516/2015/12/048}
\end{barticle}
\endbibitem

\bibitem[\protect\citeauthoryear{G\'omez-Valent
  et~al.}{2015b}]{Gomez-Valent:2014rxa}
\begin{barticle}
\bauthor{\bsnm{G\'omez-Valent}, \binits{A.}},
\bauthor{\bsnm{Sol\`{a}}, \binits{J.}},
\bauthor{\bsnm{Basilakos}, \binits{S.}}:
\bjtitle{JCAP}
\bvolume{1501},
\bfpage{004}
(\byear{2015}b).
\arxivurl{arXiv:1409.7048}.
doi:\doiurl{10.1088/1475-7516/2015/01/004}
\end{barticle}
\endbibitem

\bibitem[\protect\citeauthoryear{Grande et~al.}{2011}]{Grande:2011xf}
\begin{barticle}
\bauthor{\bsnm{Grande}, \binits{J.}},
\bauthor{\bsnm{Sol\`{a}}, \binits{J.}},
\bauthor{\bsnm{Basilakos}, \binits{S.}},
\bauthor{\bsnm{Plionis}, \binits{M.}}:
\bjtitle{JCAP}
\bvolume{1108},
\bfpage{007}
(\byear{2011}).
\arxivurl{arXiv:1103.4632}.
doi:\doiurl{10.1088/1475-7516/2011/08/007}
\end{barticle}
\endbibitem

\bibitem[\protect\citeauthoryear{Jordan}{1949}]{Jordan:1949zz}
\begin{barticle}
\bauthor{\bsnm{Jordan}, \binits{P.}}:
\bjtitle{Nature}
\bvolume{164},
\bfpage{637}
(\byear{1949}).
doi:\doiurl{10.1038/164637a0}
\end{barticle}
\endbibitem

\bibitem[\protect\citeauthoryear{Khodam-Mohammadi
  et~al.}{2014}]{Khodam-Mohammadi:2014wla}
\begin{barticle}
\bauthor{\bsnm{Khodam-Mohammadi}, \binits{A.}},
\bauthor{\bsnm{Karimkhani}, \binits{E.}},
\bauthor{\bsnm{Sheykhi}, \binits{A.}}:
\bjtitle{Int. J. Mod. Phys.}
\bvolume{D23}(\bissue{10}),
\bfpage{1450081}
(\byear{2014}).
\arxivurl{arXiv:1409.3115}.
doi:\doiurl{10.1142/S0218271814500813}
\end{barticle}
\endbibitem

\bibitem[\protect\citeauthoryear{{Khokhlov} et~al.}{1993}]{1993A&A...270..223K}
\begin{barticle}
\bauthor{\bsnm{{Khokhlov}}, \binits{A.}},
\bauthor{\bsnm{{Mueller}}, \binits{E.}},
\bauthor{\bsnm{{Hoeflich}}, \binits{P.}}:
\bjtitle{Astronomy and Astrophysics}
\bvolume{270},
\bfpage{223}
(\byear{1993})
\end{barticle}
\endbibitem

\bibitem[\protect\citeauthoryear{Li et~al.}{2015}]{Li:2015aug}
\begin{barticle}
\bauthor{\bsnm{Li}, \binits{J.-X.}},
\bauthor{\bsnm{Wu}, \binits{F.-Q.}},
\bauthor{\bsnm{Li}, \binits{Y.-C.}},
\bauthor{\bsnm{Gong}, \binits{Y.}},
\bauthor{\bsnm{Chen}, \binits{X.-L.}}:
\bjtitle{Res. Astron. Astrophys.}
\bvolume{15}(\bissue{12}),
\bfpage{2151}
(\byear{2015}).
\arxivurl{arXiv:1511.05280}.
doi:\doiurl{10.1088/1674-4527/15/12/003}
\end{barticle}
\endbibitem

\bibitem[\protect\citeauthoryear{Li et~al.}{2011}]{Li:2011sd}
\begin{barticle}
\bauthor{\bsnm{Li}, \binits{M.}},
\bauthor{\bsnm{Li}, \binits{X.-D.}},
\bauthor{\bsnm{Wang}, \binits{S.}},
\bauthor{\bsnm{Wang}, \binits{Y.}}:
\bjtitle{Commun. Theor. Phys.}
\bvolume{56},
\bfpage{525}
(\byear{2011}).
\arxivurl{arXiv:1103.5870}.
doi:\doiurl{10.1088/0253-6102/56/3/24}
\end{barticle}
\endbibitem

\bibitem[\protect\citeauthoryear{Li et~al.}{2013}]{Li:2013nwa}
\begin{barticle}
\bauthor{\bsnm{Li}, \binits{Y.-C.}},
\bauthor{\bsnm{Wu}, \binits{F.-Q.}},
\bauthor{\bsnm{Chen}, \binits{X.}}:
\bjtitle{Phys. Rev.}
\bvolume{D88},
\bfpage{084053}
(\byear{2013}).
\arxivurl{arXiv:1305.0055}.
doi:\doiurl{10.1103/PhysRevD.88.084053}
\end{barticle}
\endbibitem

\bibitem[\protect\citeauthoryear{Padmanabhan}{2003}]{Padmanabhan:2002ji}
\begin{barticle}
\bauthor{\bsnm{Padmanabhan}, \binits{T.}}:
\bjtitle{Phys. Rept.}
\bvolume{380},
\bfpage{235}
(\byear{2003}).
\arxivurl{arXiv:hep-th/0212290}.
doi:\doiurl{10.1016/S0370-1573(03)00120-0}
\end{barticle}
\endbibitem

\bibitem[\protect\citeauthoryear{Peebles and Ratra}{2003}]{Peebles:2002gy}
\begin{barticle}
\bauthor{\bsnm{Peebles}, \binits{P.J.E.}},
\bauthor{\bsnm{Ratra}, \binits{B.}}:
\bjtitle{Rev. Mod. Phys.}
\bvolume{75},
\bfpage{559}
(\byear{2003}).
\bcomment{[,592(2002)]}.
\arxivurl{arXiv:astro-ph/0207347}.
doi:\doiurl{10.1103/RevModPhys.75.559}
\end{barticle}
\endbibitem

\bibitem[\protect\citeauthoryear{Perlmutter et~al.}{1999}]{Perlmutter:1998np}
\begin{barticle}
\bauthor{\bsnm{Perlmutter}, \binits{S.}}, \betal:
\bjtitle{Astrophys. J.}
\bvolume{517},
\bfpage{565}
(\byear{1999}).
\arxivurl{arXiv:astro-ph/9812133}.
doi:\doiurl{10.1086/307221}
\end{barticle}
\endbibitem

\bibitem[\protect\citeauthoryear{Riess et~al.}{1998}]{Riess:1998cb}
\begin{barticle}
\bauthor{\bsnm{Riess}, \binits{A.G.}}, \betal:
\bjtitle{Astron. J.}
\bvolume{116},
\bfpage{1009}
(\byear{1998}).
\arxivurl{arXiv:astro-ph/9805201}.
doi:\doiurl{10.1086/300499}
\end{barticle}
\endbibitem

\bibitem[\protect\citeauthoryear{Sahni et~al.}{2014}]{Sahni:2014ooa}
\begin{barticle}
\bauthor{\bsnm{Sahni}, \binits{V.}},
\bauthor{\bsnm{Shafieloo}, \binits{A.}},
\bauthor{\bsnm{Starobinsky}, \binits{A.A.}}:
\bjtitle{Astrophys. J.}
\bvolume{793}(\bissue{2}),
\bfpage{40}
(\byear{2014}).
\arxivurl{arXiv:1406.2209}.
doi:\doiurl{10.1088/2041-8205/793/2/L40}
\end{barticle}
\endbibitem

\bibitem[\protect\citeauthoryear{Sol\`{a}}{2013}]{Sola:2013gha}
\begin{barticle}
\bauthor{\bsnm{Sol\`{a}}, \binits{J.}}:
\bjtitle{J. Phys. Conf. Ser.}
\bvolume{453},
\bfpage{012015}
(\byear{2013}).
\arxivurl{arXiv:1306.1527}.
doi:\doiurl{10.1088/1742-6596/453/1/012015}
\end{barticle}
\endbibitem

\bibitem[\protect\citeauthoryear{Sol\`{a}}{2015}]{Sola:2015csa}
\begin{barticle}
\bauthor{\bsnm{Sol\`{a}}, \binits{J.}}:
\bjtitle{Int. J. Mod. Phys.}
\bvolume{D24}(\bissue{12}),
\bfpage{1544027}
(\byear{2015}).
\arxivurl{arXiv:1505.05863}.
doi:\doiurl{10.1142/S0218271815440277}
\end{barticle}
\endbibitem

\bibitem[\protect\citeauthoryear{Sol\`{a}}{2017}]{Sola:2016vis}
\begin{bchapter}
\bauthor{\bsnm{Sol\`{a}}, \binits{J.}}:
In: \bbtitle{{Proceedings, 14th Marcel Grossmann Meeting on Recent Developments
  in Theoretical and Experimental General Relativity, Astrophysics, and
  Relativistic Field Theories (MG14) (In 4 Volumes): Rome, Italy, July 12-18,
  2015}},
vol. \bseriesno{3},
p. \bfpage{2363}
(\byear{2017}).
\arxivurl{arXiv:1601.01668}.
doi:\doiurl{10.1142/9789813226609-0276}
\end{bchapter}
\endbibitem

\bibitem[\protect\citeauthoryear{Sol\`{a}}{2018}]{sola:2018-BD}
\begin{barticle}
\bauthor{\bsnm{Sol\`{a}}, \binits{J.}}:
\bjtitle{Int.J.Mod.Phys. D}
\bvolume{D27}(\bissue{14}),
\bfpage{1847029}
(\byear{2018}).
\arxivurl{arXiv:1805.09810}.
doi:\doiurl{10.1142/S0218271818470296}
\end{barticle}
\endbibitem

\bibitem[\protect\citeauthoryear{Sol\`{a} and
  G\'omez-Valent}{2015}]{Sola:2015rra}
\begin{barticle}
\bauthor{\bsnm{Sol\`{a}}, \binits{J.}},
\bauthor{\bsnm{G\'omez-Valent}, \binits{A.}}:
\bjtitle{Int. J. Mod. Phys.}
\bvolume{D24},
\bfpage{1541003}
(\byear{2015}).
\arxivurl{arXiv:1501.03832}.
doi:\doiurl{10.1142/S0218271815410035}
\end{barticle}
\endbibitem

\bibitem[\protect\citeauthoryear{Sol\`{a} et~al.}{2017}]{Sola:2017Hdata}
\begin{barticle}
\bauthor{\bsnm{Sol\`{a}}, \binits{J.}},
\bauthor{\bsnm{G\'omez-Valent}, \binits{A.}},
\bauthor{\bsnm{Cruz~Pérez}, \binits{J.}}:
\bjtitle{The Astrophysical Journal}
\bvolume{836}(\bissue{1}),
\bfpage{43}
(\byear{2017}).
\arxivurl{arXiv:1602.02103v5}.
doi:\doiurl{10.3847/1538-4357/836/1/43}
\end{barticle}
\endbibitem

\bibitem[\protect\citeauthoryear{Sol\`{a} et~al.}{2016}]{sola-karimkhani:2016}
\begin{barticle}
\bauthor{\bsnm{Sol\`{a}}, \binits{J.}},
\bauthor{\bsnm{Karimkhani}, \binits{E.}},
\bauthor{\bsnm{Khodam-Mohammadi}, \binits{A.}}:
\bjtitle{Class. Quantum Grav. J.}
\bvolume{34}(\bissue{2}),
\bfpage{025006}
(\byear{2016}).
\arxivurl{arXiv:1609.00350}.
doi:\doiurl{10.1088/1361-6382/34/2/025006}
\end{barticle}
\endbibitem

\bibitem[\protect\citeauthoryear{Sol\`{a}~Peracaula
  et~al.}{2018}]{Sola:2016ecz}
\begin{barticle}
\bauthor{\bsnm{Sol\`{a}~Peracaula}, \binits{J.}},
\bauthor{\bparticle{de} \bsnm{Cruz~Perez}, \binits{J.}},
\bauthor{\bsnm{G\'omez-Valent}, \binits{A.}}:
\bjtitle{EPL}
\bvolume{121}(\bissue{3}),
\bfpage{39001}
(\byear{2018}).
\arxivurl{arXiv:1606.00450}.
doi:\doiurl{10.1209/0295-5075/121/39001}
\end{barticle}
\endbibitem

\bibitem[\protect\citeauthoryear{Sugiura}{1978}]{Sugiura:1978}
\begin{barticle}
\bauthor{\bsnm{Sugiura}}:
\bjtitle{Commun.Stat.}
\bvolume{A7},
\bfpage{13}
(\byear{1978}).
doi:\doiurl{10.1080/03610927808827599}
\end{barticle}
\endbibitem

\bibitem[\protect\citeauthoryear{Suzuki et~al.}{2012}]{Suzuki:2011hu}
\begin{barticle}
\bauthor{\bsnm{Suzuki}, \binits{N.}}, \betal:
\bjtitle{Astrophys. J.}
\bvolume{746},
\bfpage{85}
(\byear{2012}).
\arxivurl{arXiv:1105.3470}.
doi:\doiurl{10.1088/0004-637X/746/1/85}
\end{barticle}
\endbibitem

\bibitem[\protect\citeauthoryear{Tsujikawa et~al.}{2008}]{Tsujikawa:2008uc}
\begin{barticle}
\bauthor{\bsnm{Tsujikawa}, \binits{S.}},
\bauthor{\bsnm{Uddin}, \binits{K.}},
\bauthor{\bsnm{Mizuno}, \binits{S.}},
\bauthor{\bsnm{Tavakol}, \binits{R.}},
\bauthor{\bsnm{Yokoyama}, \binits{J.}}:
\bjtitle{Phys. Rev.}
\bvolume{D77},
\bfpage{103009}
(\byear{2008}).
\arxivurl{arXiv:0803.1106}.
doi:\doiurl{10.1103/PhysRevD.77.103009}
\end{barticle}
\endbibitem

\bibitem[\protect\citeauthoryear{Weinberg}{1989}]{Weinberg:1988cp}
\begin{barticle}
\bauthor{\bsnm{Weinberg}, \binits{S.}}:
\bjtitle{Rev. Mod. Phys.}
\bvolume{61},
\bfpage{1}
(\byear{1989}).
\bcomment{[,569(1988)]}.
doi:\doiurl{10.1103/RevModPhys.61.1}
\end{barticle}
\endbibitem

\bibitem[\protect\citeauthoryear{Wu and Chen}{2010}]{Wu:2009zb}
\begin{barticle}
\bauthor{\bsnm{Wu}, \binits{F.}},
\bauthor{\bsnm{Chen}, \binits{X.}}:
\bjtitle{Phys. Rev.}
\bvolume{D82},
\bfpage{083003}
(\byear{2010}).
\arxivurl{arXiv:0903.0385}.
doi:\doiurl{10.1103/PhysRevD.82.083003}
\end{barticle}
\endbibitem

\end{thebibliography}

\end{document}